\documentclass[10pt]{iopart}
\usepackage{bm}
\usepackage{amsfonts}

\usepackage{graphicx}
\usepackage{color}

\newcommand{\be}{\begin{equation}}
\newcommand{\ee}{\end{equation}}
\newcommand{\ba}{\begin{eqnarray}}
\newcommand{\ea}{\end{eqnarray}}
\newcommand{\bd}{\begin{displaymath}}
\newcommand{\ed}{\end{displaymath}}
\renewcommand{\vec}[1]{\mbox{\boldmath$#1$}}

\begin{document}

\title[Vorticity in Exact Model]{Study of vorticity in an exact rotating hydro model}

\author{L.P. Csernai and J.H. Inderhaug}

\address{Department of Physics and Technology, University of Bergen,
Allegaten 55, 5007 Bergen, Norway}

\ead{csernai@ift.uib.no}

\date{\today}

\begin{abstract}
We study a semianalytic exact solution of the fluid dynamical
model of heavy ion reactions, and evaluate some observable signs
of the rotation.
\end{abstract}


\maketitle

\section{Introduction}

In peripheral heavy ion collisions the
system has angular momentum.\cite{xnwang}
It has been shown in hydrodynamical computations that
the angular momentum leads to a large shear and vorticity \cite{CMW13}.
Furthermore when the Quark-Gluon Plasma
(QGP) is formed with low viscosity \cite{CKM},
interesting new phenomena may occur
like rotation \cite{hydro1}, or turbulence,
which shows up in form of a starting
Kelvin-Helmholtz instability (KHI) \cite{hydro2,WNC13}.
The deceleration of interpenetrating nuclei was observed and
analyzed early in Ref. \cite{CK85}. This leads to a rapid initial
equilibration and to the development of a compact initial system.
In peripheral collision this leads to considerable initial shear and
vorticity, as well as to an almost complete conservation of the
initial pre-collision angular momentum for the participants.

Based on Refs. \cite{CMW13,hydro2} we can extract some basic parameters
of the rotation obtained with numerical fluid dynamical model
PICR.  These parameters are extracted from model calculation
of a Pb+Pb collision
at $\sqrt{s_{NN}} = 2.76$ TeV/nucl. and impact parameter
$b=0.7 b_{max}$, with high resolution and
thus small numerical viscosity. Thus, in this collision
the KHI occurs and enhances rotation at intermediate times,
because the turbulent rotation gains energy from the
original shear flow. The turbulent rotations leads to a
rotation profile where the rotation of the external regions
lags behind the rotation of the internal zones. This is
a typical growth of the KHI.

The time dependence of some characteristic parameters of the
fluid dynamical calculation \cite{hydro2} were analysed in
Ref. \cite{CWC2014}.
It was observed that
R, the average transverse radius, Y, the longitudinal
(rotation axis directed) length  of the
participant system, $\theta$, the polar angle of the rotation
of the interior region of the system, are increasing with time.
$\dot{R}$ and $\dot{Y}$ the speeds of expansion in transverse and
axis directions are also increasing with time, while
$\omega$  the angular velocity of the
internal region of the matter during the collision is decreasing.

The initial angular momentum of the system is large,
$ L_y = - 1.05\times 10^4 \hbar$. As this is arising from the $z$
directed beam velocity, initially at the  vertical, $x$, edges the
velocity difference is large, while horizontally the rotation
starts up delayed, because this is not a solid body rotation.
Here we considered the rotation measure versus the horizontal, $z$
axis which starts up slower and reaches a maximum around 5 fm/c
after the start of the fluid dynamical evolution, i.e. around
8 fm/c after the initial touch of the nuclear surfaces.

Exact models, see e.g. Ref. \cite{Hatta}, 
provide good insight into physical phenomena.
We want to use the above mentioned fluid dynamical
calculations to test a new family
of exact rotation solutions of fireball hydrodynamics
\cite{CMW13,CsNa13}. This model offers a few possible variations,
here we chose the version 1A to test.
We use the axis labeling of Refs. \cite{CMW13,CsNa13}, so that the
the axis of the rotation is $y$ while the transverse plane of the
rotation is the $[x,z]$ plane. Thus the values extracted from the
results of the fluid dynamical model \cite{hydro2}, should take
this into account. The initial radius parameter, $R$, corresponds
to the $x$ axis in hydro, and we assume an $x,z$ symmetry in the
exact model, The rotation axis is the $y$ axis in hydro.
 The exact model assumes cylindrical
symmetry, so it cannot describe the beam directed elongation of
the system, but this is arising from the initial beam momentum,
and we intend to describe the rotation of the interior part
of the reaction plane and the rotation there.

For simplicity we also
assume that the Equation of State (EoS) is
\be
\epsilon = \kappa p \ \ \ {\rm and} \ \ \ p = nT \ ,
\ee
with a constant $\kappa$.

\section{From the Euler Equation to Scaling}

Now we calculate the equation of motion, (15) in
Ref.\cite{CsNa13}, and its solution
\be
n\,m\, (\partial_t + \vec{v}\cdot\nabla) \vec{v} = - \nabla p
\label{Euler}
\ee

 For the variables of this equation we have:
\ba
T &=&T_0\left(\frac{V_0}{V}\right)^{1/\kappa}\mathcal{T}(s)\ ,
\nonumber \\
n &=&n_0\frac{V_0}{V}\nu (s) ,
\nonumber \\
\nu(s) &=& \frac{1}{\mathcal{T}(s)}
e^{-\frac{1}{2}\int_0^s\frac{d u}{\mathcal{T}(u)}},
\label{defs}
\ea
and in addition in Ref. \cite{CsNa13} it is assumed that
the temperature and the density have time independent distributions
with respect to a scaling variable:
$$
 s = r_x^2/R^2 + r_y^2/Y^2 + r_z^2/R^2 \ .
$$
If we asume cylindrical symmetry and
use the coresponding
cylindrical coordinates instead of $(x, y, z)$,
we can use the coordinates of length
dimension, $(r_\rho, r_\varphi, r_y)$,
so that
$$
r_\rho = \rho,\ \  r_\varphi= r_\rho \varphi,\ \  r_y = y \ .
$$
These are the  "out, side, long" directions.
The characteristic values of these coordinates are then $(R, S, Y)$.
Then the scaling variables are introduced as
$$
s_\rho    =   r_\rho^2/R^2, \ \
s_\varphi =   r_\varphi^2/S^2 \ \
s_y       =   r_y^2/Y^2 \ ,
$$
where $S$ is the roll-length on the outside circumference,
 starting from $\varphi_{0} = 0$ and $S_0=0$ at $t_0$, \ \
$S = R \varphi$  and $\dot{\varphi}~=~\omega$ and this
displacement is orthogonal to the longitudinal and transverse
displacements. The internal roll-length
$r_\varphi   = \varphi\,   r_\rho $, the corresponding velocity is
$v_\varphi   = \omega\,    r_\rho $, and so
$v_\varphi^2 = \omega^2\,  r_\rho^2 $. On the other hand from the
scaling of $r_\rho$, it follows that $ r_\rho^2 = R^2\, s_\rho$.

  In case of these
scaling variables the distributions of density and temperature,
$n(s)$ and $T(t,s)$
should not depend on $s_\varphi$ or  $r_\varphi$, just on the radius and the
longitudinal coordinates. Therefore just as in Ref. \cite{CWC2014}
we introduce another scaling variable:
$$
s \equiv s_\rho + s_y\ .
$$
Our reference frame is then spanned by the
directions $(r_\rho, r_\varphi, r_y)$.
In this case due to the cylindrical symmetry the derivatives,
$\partial s/\partial r_\varphi$ vanish.
In this coordinate system the volume is $V = \pi  R^2 Y$.
\bigskip


Now following Ref. \cite{CWC2014},
for the {\bf right hand side} of Eq. (\ref{Euler}):
For the r.h.s. of this equation we have:
\ba
-\nabla p &=& -\nabla n T \nonumber\\
&=& - n_0\frac{V_0}{V}T_0\left(\frac{V_0}{V}\right)^{1/\kappa}
 \nabla e^{-\frac{1}{2}\int_0^s\frac{du}{\mathcal{T}(u)}}
\nonumber\\
&=& - n_0\frac{V_0}{V}T_0\left(\frac{V_0}{V}\right)^{1/\kappa}
     e^{-\frac{1}{2}\int_0^s\frac{du}{\mathcal{T}(u)}}
     (-\frac{1}{2})\frac{1}{\mathcal{T}(s)}\nabla s
\nonumber\\
&=&n m Q / V^\gamma
\left(\frac{r_\rho}{R^2}\vec{e}_\rho {+}\frac{r_z}{Y^2}\vec{e}_z \right)
\label{eul}
\ea
where $\gamma = 1/\kappa$ and $Q\equiv\frac{T_0 V_0^\gamma}{m}$.

\bigskip  

Using the $\rho, \varphi, y$ coordinates, the rotation
would show up as an independent orthogonal term.
However, (as discussed in the Appendix)
the closed system has no external torque, and the
internal force from the gradient of the pressure is
radial, which does not contribute to tangential acceleration.
The change of the angular velocity arises from the angular momentum
conservation in the closed system as a constraint, so we do not have
to derive additional dynamical equations to describe the evolution
of the rotation.

Now for the {\bf left hand side} of Eq. (\ref{Euler}),
the velocity field scales as
\be
\vec{v} =
v_\rho \vec{e}_\rho {-} v_\varphi \vec{e}_\varphi {+} v_z \vec{e}_z =
\frac{\dot{R}}{R} r_\rho \vec{e}_\rho {-}
\omega r_\rho \vec{e}_\varphi {+}
\frac{\dot{Y}}{Y} r_y \vec{e}_y \ .
\label{velocity}
\ee
We first calculate the time derivatives for the components.
(See e.g. \cite{Stoecker_handbook}):
\ba
\partial_t v_\rho &=& \left[ \left(
\frac{\ddot{R}}{R}{-}\frac{\dot{R}^2}{R^2}\right) - \omega^2 \right]  r_\rho ,
\nonumber \\
\partial_t v_\varphi &=&
- \omega \frac{\dot{R}}{R} r_\rho , \ \ \
\partial_t v_z =
\left[ \frac{\ddot{Y}}{Y}{-}\frac{\dot{Y}^2}{Y^2}\right]  r_y .
\label{partialtvx}
\ea
The other term of the comoving derivative includes
\ba
\vec{v}\cdot\nabla =
v_\rho\frac{\partial}{\partial r_\rho}+
v_\varphi\frac{\partial}{\partial r_\varphi}+
v_y\frac{\partial}{\partial r_y}
\ea
and this term gives:
\be
(\vec{v}\cdot\nabla) \vec{v} =
\frac{\dot{R}^2}{R^2}r_\rho     \vec{e}_\rho +
\omega \frac{\dot{R}}{R} r_\rho \vec{e}_\varphi +
\frac{\dot{Y}^2}{Y^2}r_y \vec{e}_y
\label{vDeltvx}
\ee
By adding Eq. (\ref{partialtvx}) and (\ref{vDeltvx}) we get:
\ba
mn(\partial_t{+}\vec{v}\cdot\nabla)v_\rho
 &=& mn\left[\left( \ddot{R}/R\right) - \omega^2 \right] r_\rho \ ,
\nonumber \\
mn(\partial_t{+}\vec{v}\cdot\nabla)v_y
 &=& mn\left( \ddot{Y}/Y\right) r_y \ .
\ea

As a consequence
the {\bf equality} of the right hand side and left hand side
of the Euler equation (\ref{Euler}) leads
to the ordinary differential equations.
Multiplying the two non-vanishing equations with $R^2$ and $Y^2$ respectively
yields:
\ba
R\ddot{R} - W/R^2 &=& Y\ddot{Y} =
 \frac{Q}{(\pi R^2Y)^{\gamma}} \ ,
\label{e-2s}
\ea
where $W~\equiv~\omega_0^2~R_0^4$.
From the angular momentum conservation
$\omega=\omega_0R_0^2/R^2$, and the rotational
term, $R^2\omega^2$ that appears in the equation,
takes the form  $W / R^2$.

Due to the EoS the pressure is proportional to the
baryon density $n$, just as the r.h.s. of the Euler equation,
therefore the equation of motion does not depend on $n$ or $n_0$.

\bigskip

\section{Conservation Laws}
\bigskip

Following Ref. \cite{CWC2014},
we want to calculate the total energy of the whole system,
then we should integrate it for the whole volume, $V$.
Thus, not only the scaling of
$\vec{v} = ( v_\rho, v_\varphi, v_z)$ but also the particle density
distribution, $n(s)$ will be considered.
\smallskip

At the {\bf surface}  the rotational energy  is
$\mathcal{E}_{Side} \equiv \frac{1}{2} m \dot{S}^2 =
 \frac{1}{2} m R^2 \omega^2$,
and if we express $\omega$ via $\omega_0$ by the relation
$\omega = \omega_0 R_0^2 / R^2$, then
$\mathcal{E}_{Side} = W / R^2$, as before.
The expansion energy at the surface is
$\mathcal{E}_{Out} \equiv \frac{1}{2} m \dot{R}^2$, and for the
longitudinal direction we have,
$\mathcal{E}_{Long} \equiv \frac{1}{2} m \dot{Y}^2$.
\smallskip

We can calculate the radial and longitudinal
expansion velocities and the corresponding kinetic energies, and also
the kinetic energy of the rotation.
In the evaluation of the internal and kinetic
energies, the radial and longitudinal density profiles of the
system are taken into account.

Let us assume that the temperature
profile is flat, and thus that the density profiles are Gaussian and
separable.
Further if we assume that the system size is finite,
then the scaling variables, $s_\rho$ and $s_Y$, may extend from
0 to 1. In this case at the external boundary we have to apply the
necessary boundary conditions so that the solution of the Euler
equation (\ref{Euler}) remains valid.
With this approximation we calculated the different integrated energies
(and shown in Appendices \ref{A1}-\ref{A4}).

Summing up the kinetic energies yields
\be
E_{kin} = \frac{1}{2} m N_B \left(
\alpha^2 \dot{R}^2 +  \alpha^2 R^2 \omega^2 + \beta^2 \dot{Y}^2 \right) \ ,
\label{Etot-1}
\ee
where in case of finite extent of the system\\
$
\alpha^2\! \equiv \!
4\sqrt{2}\,C_n I_B({\small \frac{1}{2}}s_{yM})
               I_C({\small \frac{1}{2}}s_{\rho M})\,
{\rm and}\,
$\\
$
\beta^2\! \equiv\!
4\sqrt{2}\,C_n I_A({\small \frac{1}{2}}s_{\rho M})
               I_D({\small \frac{1}{2}}s_{yM})$
where\\
$C_n = 1 \left/ \left[  2\sqrt{2}\, I_A(s_{\rho M}/2)\,
                               I_B(s_{yM}/2) \right] \right.$,
see Ref. \cite{IGF} 			
\footnote{
$I_A(u) = 1 - \exp(-u)$,
$I_B(u) = \sqrt{\pi}\, \Phi(\sqrt{u}\,)$,
$I_C(u) = 1 - (1+u) \exp(-u)$,
$I_D(u) = \frac{\sqrt{\pi}}{2} \Phi(\sqrt{u}) - \sqrt{u} e^{-u}$,
where
$\Phi(u)= {\rm erf}(u) \equiv \frac{2}{\sqrt{\pi}} \int_0^u \exp(-x^2)\,dx.$
}
(in terms of the integrals evaluated in
Appendices \ref{A1}-\ref{A4} ).

Here with $(s_{\rho M}=(s_{yM}=1$, $\alpha^2$ and $\beta^2$ are clearly time independent, because
they depend on the scaling variable only, and we get the values
$\alpha^2 = 0.4585$ and
$\beta^2 = 0.2911$.

Alternatively one can assume that the system size is infinite so that the
scaling variables range from 0 to $\infty$. In this case the Radius, and
Length parameters, $R$ and $Y$, are considered as the width
of the Gaussian scaling distribution. Thus the parameters will be
\be
I_A=I_B=1,\ I_C=2I_D=\sqrt{\pi}
\label{alternate}
\ee
and consequently
$\alpha^2 = 2.0$ and
$\beta^2 = 1.0$. In the present case we follow this configuration.

If we divide this result by the conserved baryon charge,
$N_B$, we will get
\be
\frac{E_{kin}}{N_B} =  \frac{1}{2} m  \left[
 \alpha^2 \left( \dot{R}^2 {+} R^2 \omega^2 \right)
 +   \beta^2\, \dot{Y}^2 \right]\ ,
\ee

Based on the EoS, $\epsilon = \kappa p = \kappa n T$, we can
calculate the compression energy based on the
density profiles of $n(s)$ and $\epsilon(s) = \kappa\, n(s) T(s)$.

Let us make the same simplifying assumptions on the density profiles
as we did earlier.
Now we will have the same density profile, normalized to $N_B$, for
the volume integrated internal energy and the net baryon charge:

\ba
E_{int} &=& \kappa  \int p dV = \kappa \int n T dV
\nonumber \\
&=& \kappa N_B T_0 (V_0/V)^\gamma \, C_n \,
  \frac{1}{V}\ \pi R^2\!\! \int_0^1 Y\!\! \int_{0}^1\!\!\!
\nu(s)\ ds_\rho\, \frac{ds_z}{\sqrt{s_z}}
\nonumber \\
 &=&  \kappa  N_B T_0 (V_0/V)^\gamma  \ = \
\kappa m Q \frac{1}{(\pi R^2Y)^\gamma} \ ,
\ea
where $C_n$ is the normalization constant.

\section{Reduction to a Single Differential Equation}
\bigskip

Following the method of Ref. \cite{Akkelin01} ,
we  study the following combination of variables:
\ba
{\cal F} &=& {\small \frac{1}{2}} \partial_t^2
\left( \alpha^2 R^2 +  \beta^2 Y^2 \right)
\nonumber\\
 &=& \partial_t \left(
 \alpha^2 R \dot{R}  + \beta^2 Y \dot{Y} \right)
\nonumber\\
 &=& \alpha^2 \dot{R}^2  + \beta^2 \dot{Y}^2 +
\alpha^2 R \ddot{R}  + \beta^2 Y \ddot{Y}  ,
\label{F16}
\ea
where we used the notation
$
\partial_t = \frac{\partial}{\partial t} \ \ {\rm and } \ \
\partial_t^2 = \frac{\partial^2}{\partial t^2}.
$
We can replace the last two terms,
$ \alpha^2 R \ddot{R},\  \beta^2 Y \ddot{Y}$, by
using Eqs. (\ref{e-2s}), i.e. we use the Euler Eq. (\ref{Euler}). Then
we obtain:
\be
 {\cal F} =
  \alpha^2 \dot{R}^2 + \beta^2 \dot{Y}^2 + \alpha^2 \frac{W}{R^2}
+ (\alpha^2{+}\beta^2) \frac{Q}{(\pi R^2Y)^\gamma} ,
\label{EC1}
\ee

At the same time from the {\bf energy conservation},
$E_{tot} = E_{kin} + E_{int}$, we get that
\be
\frac{E_{tot} }{N_B \, m} = \frac{1}{2}
\left[
 \alpha^2 \dot{R}^2 {+} \beta^2 \dot{Y}^2 {+} \alpha^2 \frac{W}{R^2}
{+} \frac{2 \kappa Q}{(\pi R^2Y)^\gamma}
\right] ,
\label{EC2}
\ee
where we used the EoS and the parameter $\kappa$ now appears
in the expression of the energy.
If our EoS is such that
\be
     \kappa = \frac{1}{2} (\alpha^2{+}\beta^2) =\frac{3}{2},
\ee
then $ {\cal F} = 2 E_{tot} / (N_B \, m ) = $ const., and in the same type of calculation
as in Ref. \cite{Akkelin01}, we can introduce
\be
U^2(t) \equiv  \alpha^2 R^2(t) +  \beta^2 Y^2(t) ,
\label{eu}
\ee
which satisfies
\be
 \partial_t^2 \left( \alpha^2 R^2 + \beta^2 Y^2 \right) =
 \partial_t^2 U^2(t) = 2 {\cal F}  \ .
\label{u21}
\ee
Thus, the solution of Eq. (\ref{u21}), we can be parameterized as:
\be
U^2(t) = A (t-t_0)^2 + B (t-t_0) + C \ ,
\ee
where
\ba
A&=& \alpha^2 \dot{R}_0^2+\beta^2\dot{Y}_0^2+ \alpha^2 W/ R_0^2 +
(\alpha^2{+}\beta^2) \frac{T_0}{m}
\nonumber \\
B&=& 2\alpha^2 R_0 \dot{R}_0 + 2\beta^2 Y_0 \dot{Y}_0
\phantom{\frac{T_0}{m}}
\nonumber \\
C&=&  \alpha^2 R_0^2 +  \beta^2 Y_0^2     \phantom{\frac{T_0}{m}}\ .
\ea

Due to the difficulties described in Appendix \ref{A5}, we cannot use the
method described in  \cite{Akkelin01}. Instead
let us take one of the Euler equations from Eq. (\ref{e-2s}),
\be
\ddot{Y} =  \frac{Q}{Y (\pi R^2 Y)^{\gamma}} \ ,
\label{e-4s}
\ee
and express $R^2$ in terms of $U^2(t)$ which is known
based on the energy conservation:
\be
R^2(t) = (U^2(t) - \beta^2 Y^2)/ \alpha^2 \ ,
\label{er}
\ee
and this will lead to the second order differential equation
for $Y(t)$:
\be
\ddot{Y} =
\frac  {\alpha^{2\gamma}\, Q}
{ Y \left[ \pi Y (U^2(t) {-} \beta^2 Y^2)\right]^{\gamma}}
= f(Y,t)\ ,
\label{e-5s}
\ee
which can be solved \\
Then $R(t)$ and $\dot{R}(t)$ are  given
by Eqs. (\ref{er}) and (\ref{EC1}) respectively.

\begin{table}[h]
\begin{center}	
\begin{tabular}{cccccccc} \hline\hline \phantom{\Large $^|_|$}
$t$  &  $Y$ & $\dot{Y}$ & $R$ & $\dot{R}$ & $\omega$ \\
(fm/c)&\ \ (fm)\ \ &\ \ (c)\ \ \ &\ \ \ (fm)\ \ &\ \ (c)\ \ \ &\ (c/fm)\ \\
\hline
 0.0 & 4.000 & 0.400 &  2.500 &  0.250 &   0.150\\
 1.0 & 4.440 & 0.469 &  2.859 &  0.704 &   0.115\\
 2.0 & 4.922 & 0.490 &  3.405 &  0.834 &   0.081\\
 3.0 & 5.415 & 0.495 &  4.079 &  0.877 &   0.056 \\
 4.0 & 5.912 & 0.497 &  4.833 &  0.894 &   0.040 \\
 5.0 & 6.409 & 0.497 &  5.636 &  0.902 &   0.030 \\
 6.0 & 6.906 & 0.498 &  6.469 &  0.906 &   0.022 \\
 7.0 & 7.404 & 0.498 &  7.322 &  0.909 &   0.017 \\
 8.0 & 7.901 & 0.498 &  8.190 &  0.911 &   0.014 \\
\hline
\end{tabular}
\end{center}
\caption{
Time dependence of characteristic parameters of the exact
fluid dynamical model \cite{CWC2014}.
R is the transverse radius, Y is the
(rotation axis directed) length  of the
system, $\dot{R} ,\ \dot{Y}$ are the speed of expansion in transverse and
axis directions, and $\omega$ is the angular velocity of the
matter.
}
\label{t2}
\end{table}

The derivatives,
$\dot{R}(t_0)$ and $\dot{Y}(t_0)$ in this exact model do not
equal the ones obtained from the fluid dynamical model, because in the
more realistic fluid dynamical model the density and velocity profiles
do not agree with
the exact model's assumptions. Also initially in the realistic
fluid dynamical model the angular momentum increases in the
region due to the developing turbulence, while in the exact model
the angular velocity is monotonously decreasing
due to the scaling expansion.

\begin{figure}[ht]  
\begin{center}
\resizebox{0.9\columnwidth}{!}
{\includegraphics{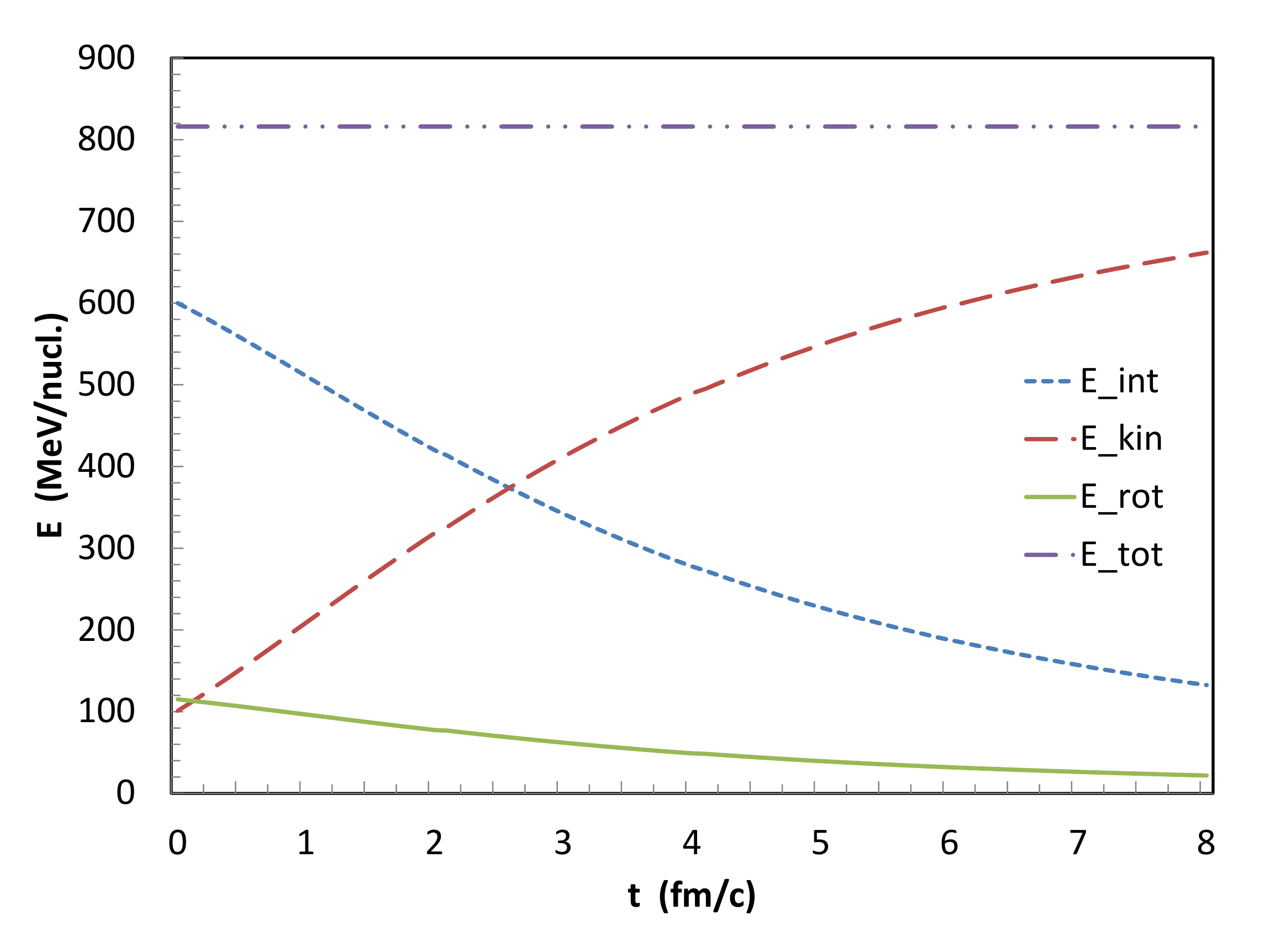}}
\caption{
(Color online) The time dependence of the kinetic energy
of the expansion, $E_{kin}$,
the internal energy, $E_{int}$,
the rotational energy, $E_{rot}$, and
the total energy, $E_{tot}$
per nucleon in the exact model
with the initial conditions
$R_0$ = 3.5 fm,  $Y_0$ = 5.0 fm,
$\dot R_0$ = 0.25 c,  $\dot Y_0$ = 0.30 c,
$\omega_0$ = 0.1 c/fm,
$\kappa = 3/2$,  $T_0$ = 400 MeV.
For this configuration $E_{tot}=816$ MeV/nucl.
The kinetic energy
of the expansion is increasing, at the cost of the decreasing
internal energy and the slower decreasing rotational
energy.
The rotational energy is decreasing to the half of the
initial one in 3.3 fm/c.
}
\label{F1-E-vs-t}
\end{center}
\end{figure}

The Runge Kutta \cite{RungeKutta} method was used to solve this
differential equation. We chose the constants, $Q$ and $W$, as well
as the initial conditions for $R$ and $Y$.

Based on the fluid dynamical
model calculation results  we chose the
parameters:
$
T_0 = 250\  {\rm MeV},\
m  =  939.57 {\rm MeV}\
\omega_0 =0.15$ c/fm.
For the internal region we take the initial radius parameters as
$
R_0 = 2.5 {\rm fm\ and\ }
\dot{R} = 0.25 {\rm c}
$,
and we disregard the larger extension
in the beam direction, because our model is cylindrically
symmetric and because the beam directed large elongation is a consequence
of the initial beam directed momentum excess.
In this exact model the rotation axis, denoted by $Y$, corresponds to the
out of plane, $y$ direction in the fluid dynamical model
(and not to the beam direction!). Due to the eccentricity at finite
impact parameters, with an almond shape profile, the initial out of plane
size is larger the in plane transverse size, so we chose initially
$
Y_0 = 4.0  {\rm fm\ and\ }
\dot{Y} = 0.4 {\rm c}
$
just as in Ref. \cite{CWC2014}. (Table \ref{t2})

\begin{figure}[hb]  
\begin{center}
\resizebox{0.49\columnwidth}{!}
{\includegraphics{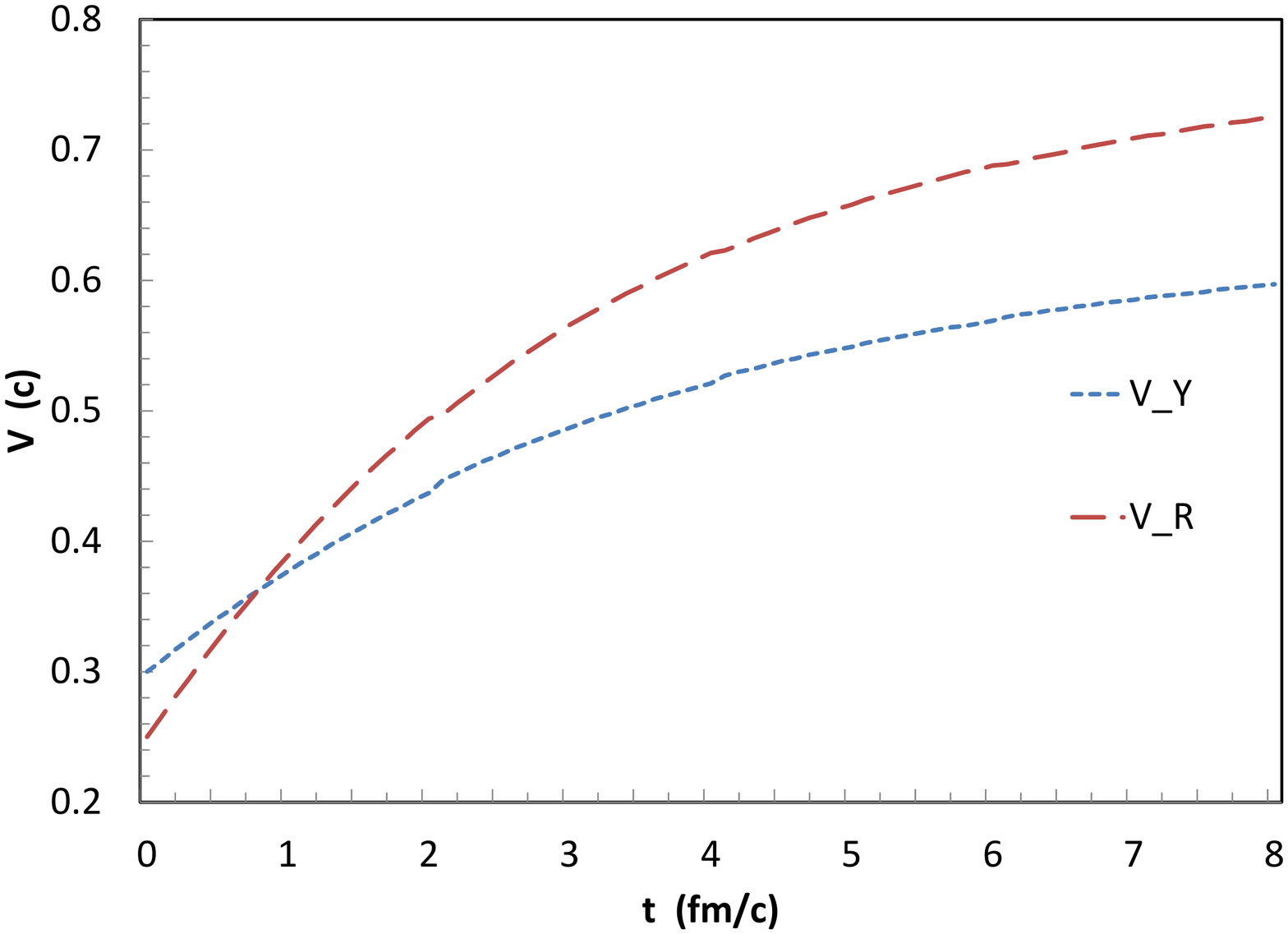}}
\hfill
\resizebox{0.49\columnwidth}{!}
{\includegraphics{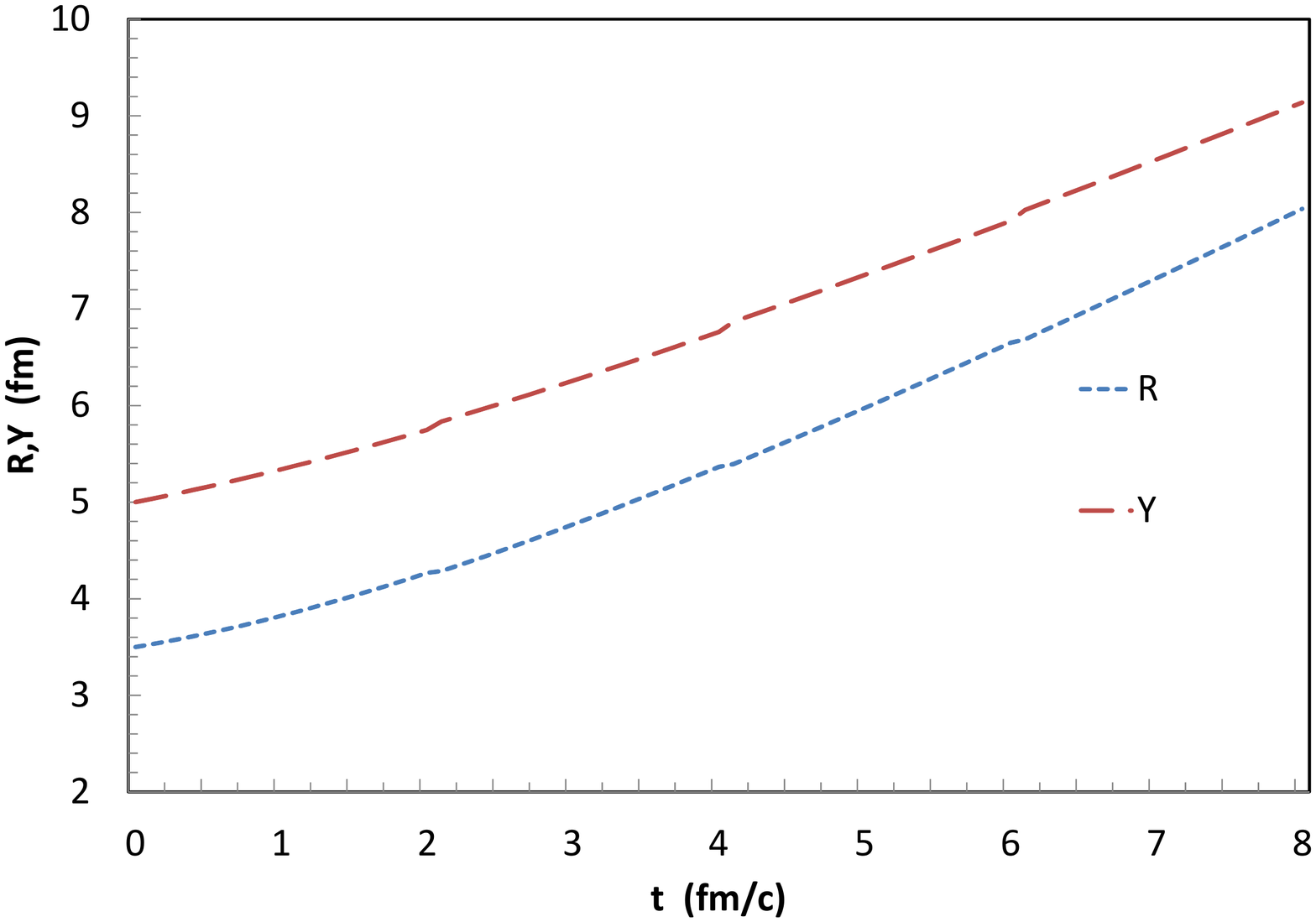}}
\caption{
(Color online)
Left: The time dependence of the velocity of expansion in the
transverse radial direction, $v_R$ and in the direction of
the axis of the rotation, $v_Y$ for the configuration shown in Fig.
\ref{F1-E-vs-t}. The expansion velocity
is increasing in both directions. While in the axis direction
the velocity increases from 0.3 c to 0.6 c in 8 fm/c time, the
radial expansion increases faster, in part due to the centrifugal
force from the rotation.
Right: The time dependence of the Radial, $R$, and axis directed, $Y$,
size of the expanding system.
As the $Y$ directed velocity is initially larger its change
is relatively smaller.
}
\label{F1-v-vs-t}
\end{center}
\end{figure}

As the exact solution is able to describe the monotonic expansion,
and so the steady decrease of the rotation, we start from a higher
initial angular velocity than shown by the fluid dynamical model, PICR,
as the angular velocity, measured versus the horizontal plane,
starts from zero.

Applying these initial parameters the exact model yields a dynamical development
shown in Table \ref{t2}. According to expectations the radius, $R$, and the
axis directed size, $Y$, are increasing, the angular velocity, $\omega$
decreases, The total energy is conserved, while the kinetic
energy of expansion is increasing, and that of the rotation and
internal energy are decreasing.
See Fig. \ref{F1-E-vs-t}.

The change of the expansion velocity, $v_R=\dot{R}$,  is shown in
Fig.  \ref{F1-v-vs-t} left.
The more rapid velocity change arises partly from the centrifugal
acceleration of the rotation, but also from the fact that the
initially smaller transverse size increases faster in the direction
of equal sizes in both directions. See Fig.  \ref{F1-v-vs-t} right.

The study of the rotation in an infinite system is. on the other hand,
problematic as we assume solid body rotation (i.e.  the
angular momentum applies to the whole infinite system). So the
applicability of this infinite model to a heavy ion reaction
is highly approximate, and the external tails should be disregarded.

Other finite scaling expansion profiles can also be studied, based on
the given examples, and these may fit more detailed fluid dynamical
models better.

\section{The vorticity}

In the usual convention in heavy ion physics, the beam axis is
the $z$ axis, the impact parameter vector, $\vec b$, points in the
$x$ direction, and the projectile is at positive $x$ and moves in the
positive $z$ direction. Thus the rotation axis is the $y$ axis,
this is the axis of the cylindrical symmetry of the rotating
exact model system we discussed above. The reaction plane $x,z$ is spanned by the
cylindrical coordinates $r, \varphi$ in the discussion above.

\begin{figure}[ht]  
\begin{center}
\resizebox{0.485\columnwidth}{!}
{\includegraphics{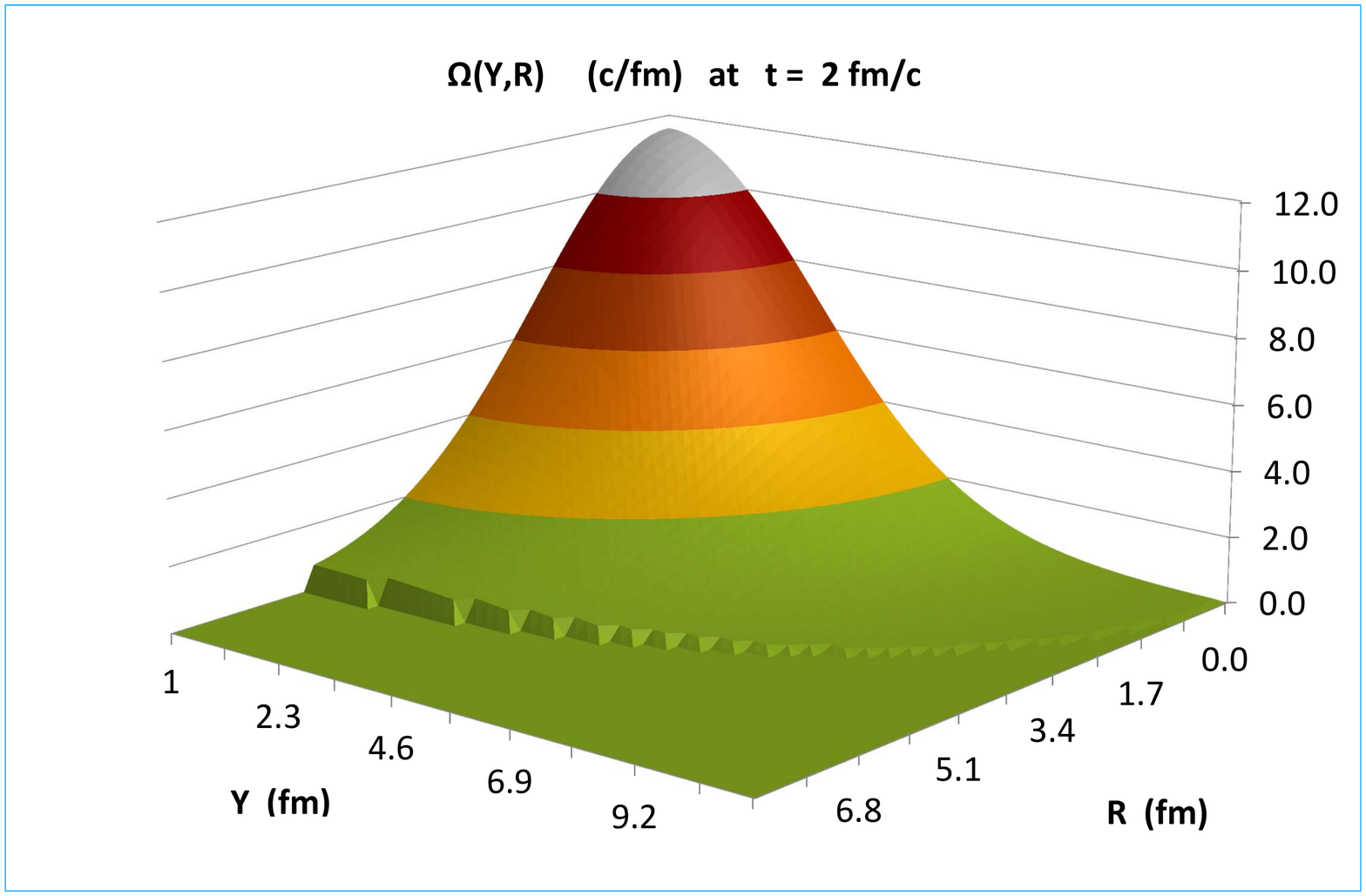}}
\hfill
\resizebox{0.485\columnwidth}{!}
{\includegraphics{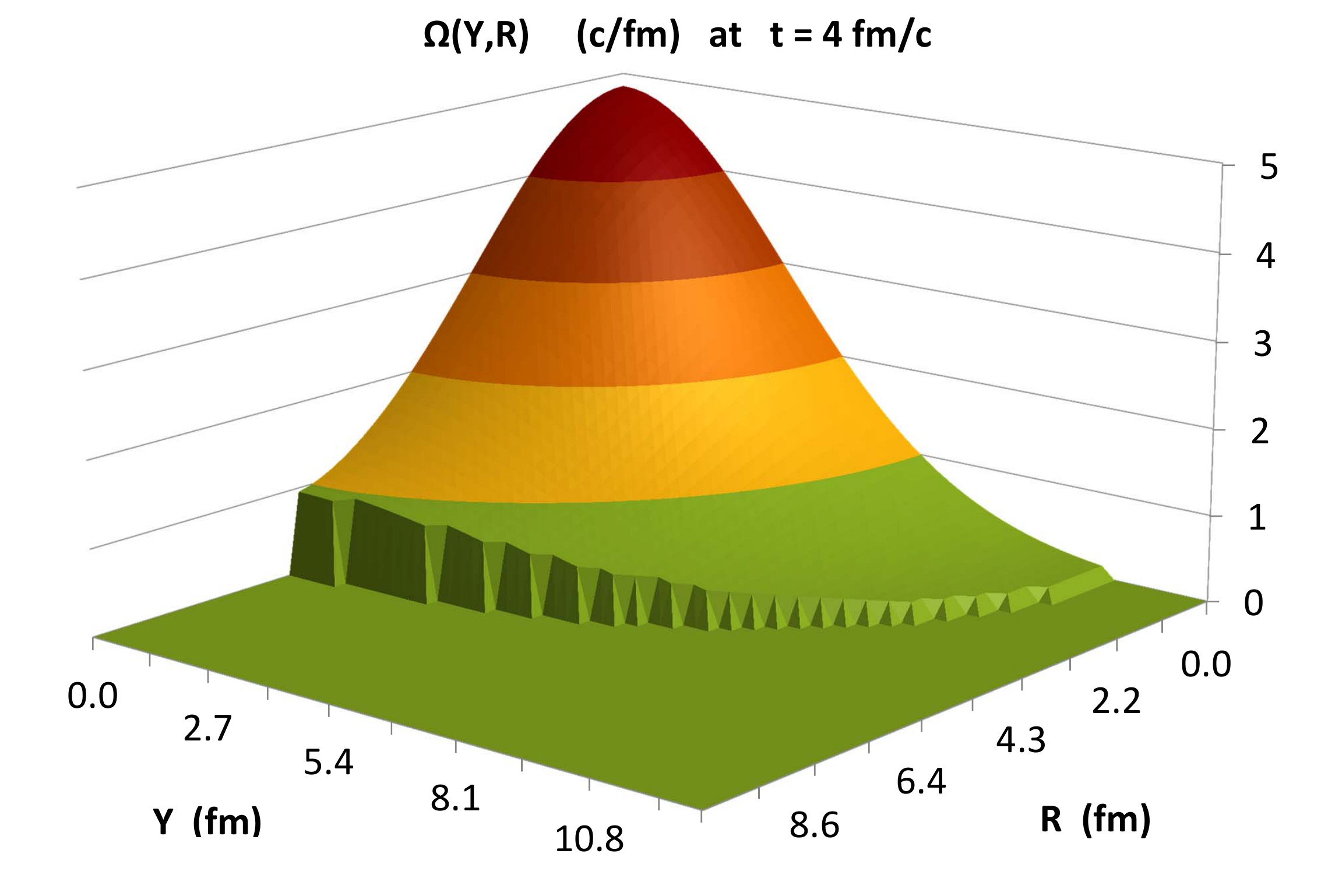}}\\
\resizebox{0.485\columnwidth}{!}
{\includegraphics{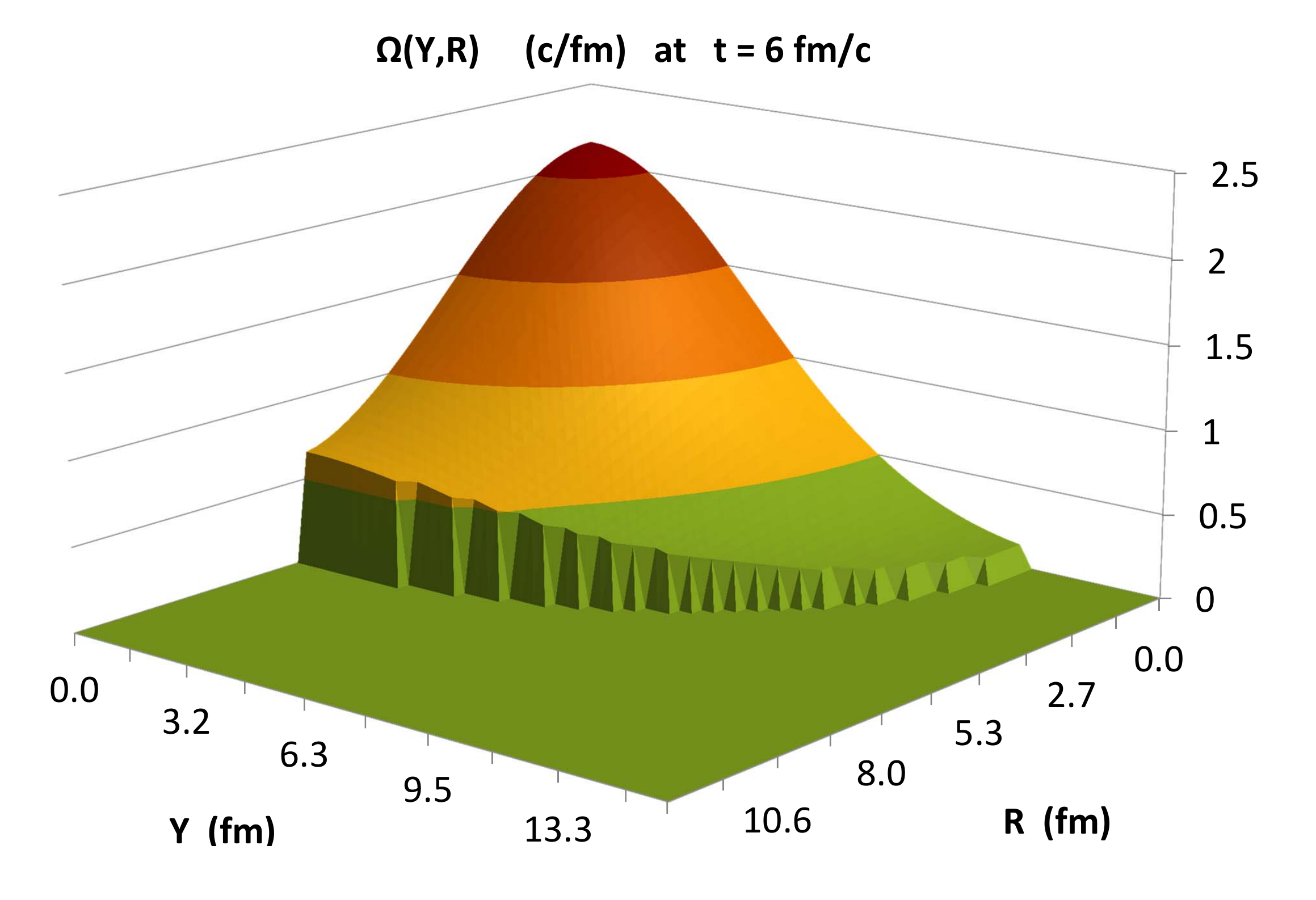}}
\hfill
\resizebox{0.485\columnwidth}{!}
{\includegraphics{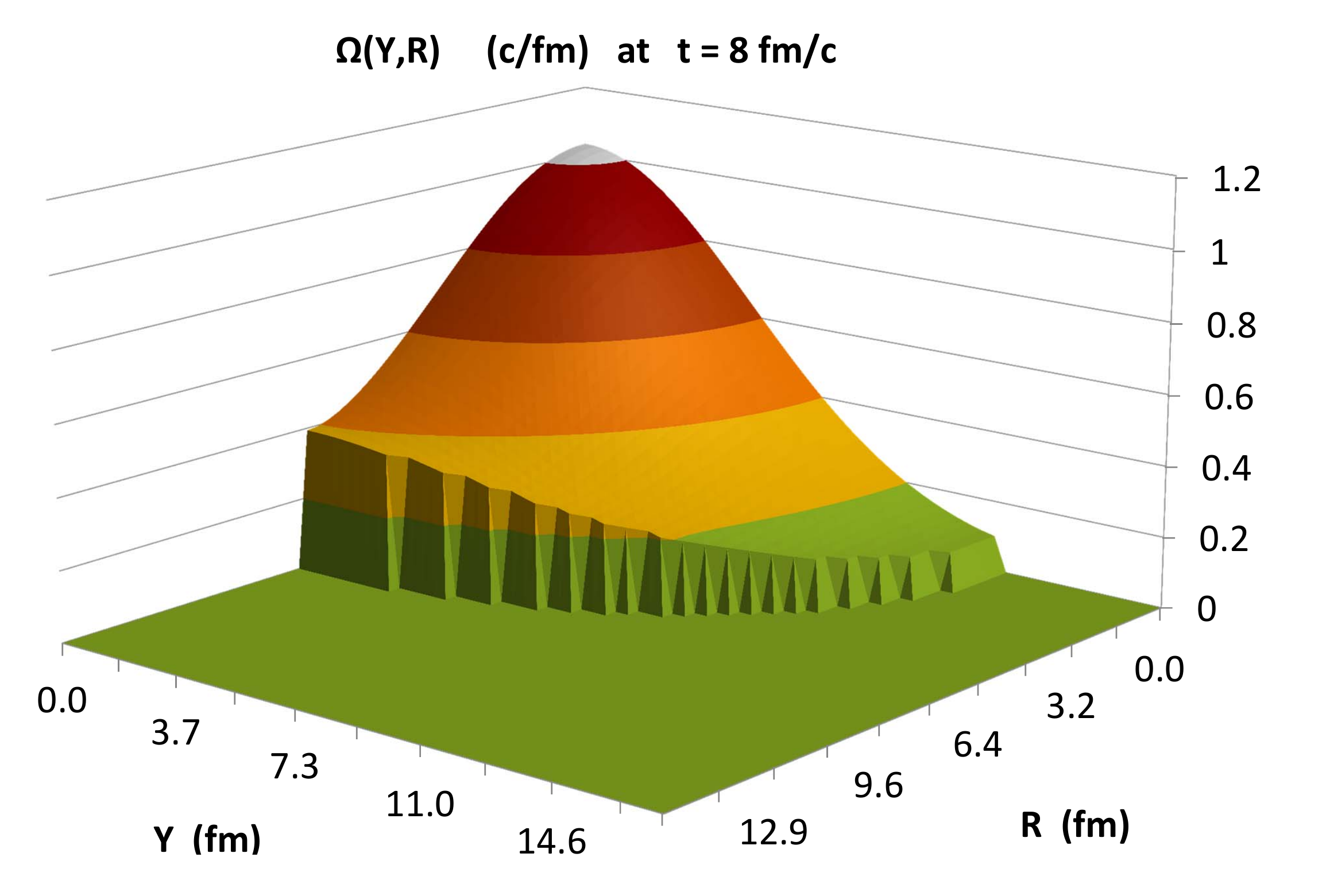}}
\caption{
(Color online)
The energy weighted vorticity in the classical rotating exact model
with Gaussian density profiles, with an EoS of $\kappa = 3/2$,
and with initial parameters:
3.5 fm mean radius,
5.0 fm mean length,
0.1 c/fm angular velocity,
0.25 c radial velocity,
0.3 c axis directed velocity ($v_{y}$).
The initial temperature of the matter is $T=400$ MeV.
The figures
show the configuration at different times.
At
$t = 2$ fm/c,
the mean radial (longitudinal) sizes and speeds are
4.27 fm (5.75 fm) and 0.494 c (0.437 c), and the
angular velocity is 0.07 c/fm.
The $v=c$-boundary is at
$Y_{max} = 11.0$ fm and $R_{max}= 6.3$ fm.
At
$t = 4$ fm/c,
the mean radial (longitudinal) sizes and speeds are
5.67 fm (6.76 fm) and 0.62 c (0.52 c), and
$\omega=0.04$ c/fm.
The $v=c$-boundary is at
$Y_{max} = 7.1 $ fm and $R_{max}= 11.1$ fm.
At
$t = 6$ fm/c,
the mean radial (longitudinal) sizes and speeds are
6.65 fm (7.91 fm) and 0.69 c (0.57 c), and $\omega= 0.0$ c/fm.
The $v=c$-boundary is at
$Y_{max} = 12.0$ fm and $R_{max}= 7.98$ fm.
At
$t = 8$ fm/c,
the mean radial (longitudinal) sizes and speeds are
8.04 fm (9.14 fm) and 0.73 c (0.60 c), and $\omega=0.02$ c/fm.
The $v=c$-boundary is at
$Y_{max} = 13.16$ fm and $R_{max}= 9.32$ fm.
}
\label{F1-t}
\end{center}
\end{figure}

For rotation around the y-axis the vorticity is defined
in terms of the velocity field as
$\omega_y=(\partial_z v_x-\partial_x v_z)/2$.  We use the
conventions of the exact model here, so that we chose the rotation axis to be
the $y$ axis and the plane of the rotation is the $x,z$ plane, which
corresponds to the reaction plane. We assume that the rotating
system is symmetric, so we introduce cylindrical coordinates around
the rotation axis.

\begin{figure}[ht]  
\begin{center}
\resizebox{0.9\columnwidth}{!}
{\includegraphics{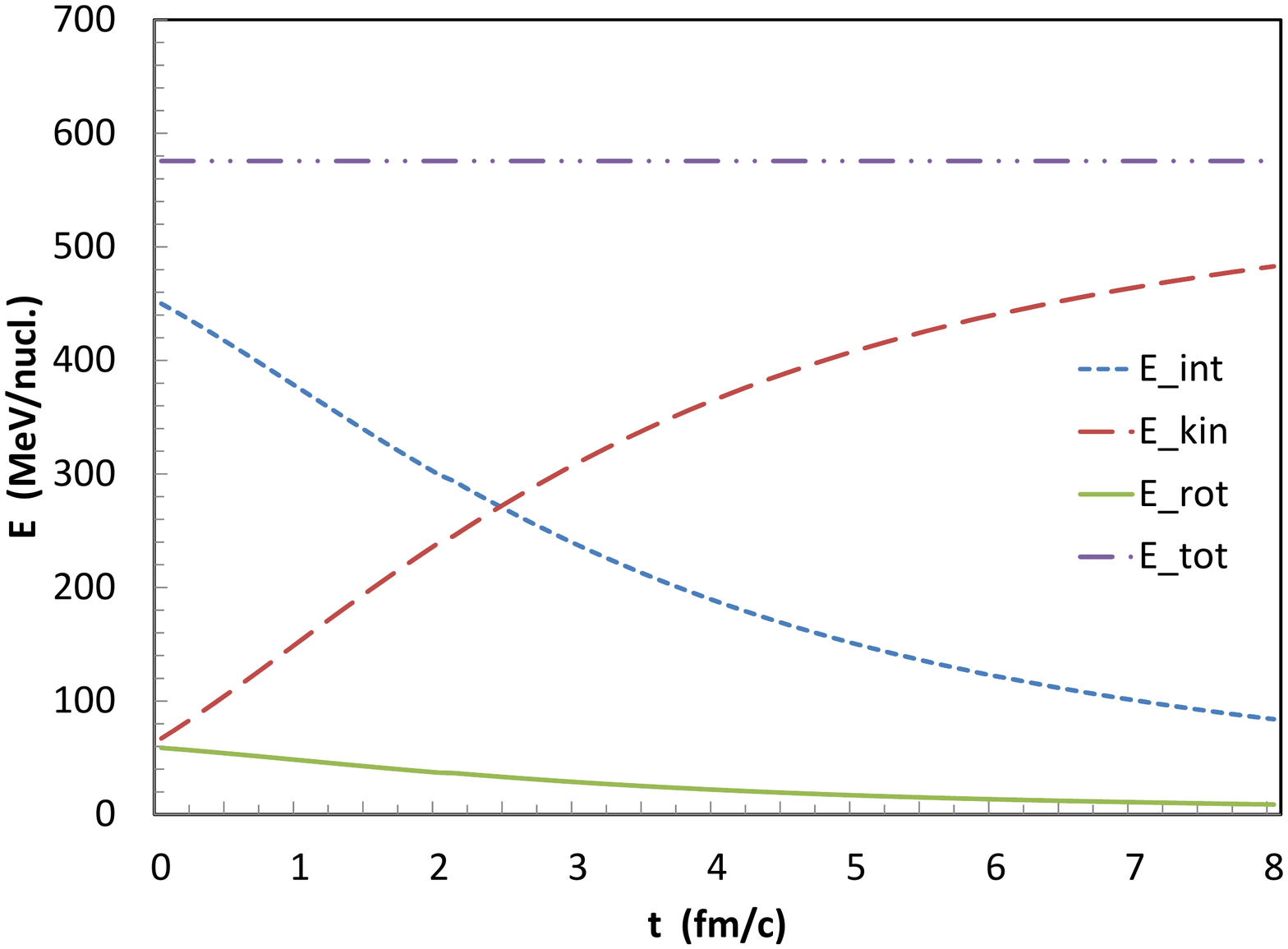}}
\caption{
(Color online) The time dependence of the kinetic energy
of the expansion, $E_{kin}$,
the internal energy, $E_{int}$,
the rotational energy, $E_{rot}$, and
the total energy, $E_{tot}$
per nucleon in the exact model
with the initial conditions
$R_0$ = 2.5 fm,  $Y_0$ = 4.0 fm,
$\dot R_0$ = 0.20 c,  $\dot Y_0$ = 0.25 c,
$\omega_0$ = 0.1 c/fm,
$\kappa = 3/2$,  $T_0$ = 300 MeV.
For this configuration $E_{tot}=576$ MeV/nucl. The kinetic energy
of the expansion is increasing, at the cost of the decreasing
internal energy and the slower decreasing rotational
energy.
The rotational energy is decreasing to the half of the
initial one in 2.9 fm/c.
}
\label{F1-E-vs2-t}
\end{center}
\end{figure}

 For this configuration in
cylindrical coordinates the vorticity is .

\ba
\vec \omega &\equiv& {\rm rot} \vec{v}
\nonumber \\
&=& \left(\frac{1}{\rho}\frac{\partial v_y}{\partial\varphi}-
    \frac{\partial v_{\varphi}}{\partial y}\right) \vec{e}_{\rho}+
    \left(\frac{\partial v_\rho}{\partial y}-
    \frac{\partial v_y}{\partial \rho}\right) \vec{e}_{\varphi} +
    \left(\frac{1}{\rho}\frac{\partial (\rho v_\varphi )}{\partial\rho}-
    \frac{1}{\rho}\frac{\partial v_{\rho}}{\partial \varphi}\right) \vec{e}_y
\nonumber \\
&=& -2\omega \vec{e}_y \, .
\ea
At the last step we use Eq. (\ref{velocity}), where
$v_y$ does not depend on $\varphi$,
$v_\varphi$ does not depend on $y$,
$v_\rho$ does not depend on $y$,
$v_y$ does not depend on $\rho$ and
$v_\rho$ does not depend on $\varphi$,
thus only one term contributes to the vorticity, which
is directed in the direction of $\vec e_y$.

Thus the vorticity in this model is spatially homogeneous, and depends
on the time only, $\omega = \omega(t)$. However, from the point
of view of observations, it is important what amount of energy or mass
is representing a given fluid element with the given vorticity.
In the solution we presented here we assumed a uniform temperature, which
led to a gaussian density and energy profile,
$\epsilon(\rho) = \kappa T n(\rho)$.

Following reference \cite{CMW13}, we define an
energy-density-weighted, average vorticity as
\be
\Omega_{zx} \equiv w(r_\rho,r_\phi,r_y) \, \omega
\ee
so that this weighting does not change the average circulation
of the layer, i.e., the sum of the average of the weights over all
fluid elements is unity, $ \langle w(z, x)\rangle = 1$.
This weighting does
not change the average vorticity value of the set; just the cells
will have larger weight with more energy content.

Let us fist calculate the internal energy for a finite system:
\be
E_{int} =
\int_{-aY}^{+aY} \int_0^{bR} \epsilon(r_\rho,r_y)\
2 \pi r_\rho dr_\rho dr_y \ ,
\nonumber \\
\ee
while the the energy at a given radius (at a given time) is
$ \epsilon(r_\rho,r_y) $.
Thus the weight density will be
\be
w(r_\rho,r_\phi,r_y) = \frac{T^{00}(r_\rho,r_y)}{E_{tot}/V} \ ,
\ee
where
$$
T^{00} = \epsilon \left[ (1+1/\kappa)\ \gamma^2 - 1/\kappa \right]
 \ \ \ \  {\rm and}
$$
\ba
\epsilon(r_\rho,r_y)
&=& \kappa n_0 \frac{V_0}{V}
T_0 \left(\frac{V_0}{V}\right)^{1/\kappa}\mathcal{T}(s)e^{-\frac{1}{2}\int_0^s
\frac{du}{\mathcal{T}(u)}}
\nonumber \\
&=& \kappa T_0 n_0  \left(\frac{V_0}{V}\right)^{1+1/\kappa}  e^{-s_{y}/2} e^{-s_{\rho}/2} ,
\ea
while
$
\gamma^2 = \left[ 1 - \left(\frac{\dot{R}}{R}r_\rho \right)^2
                    - \left(\omega r_\rho \right)^2
                    - \left(\frac{\dot{Y}}{Y}r_y \right)^2  \right]^{-1}
$.
Here we assumed that $\mathcal{T}(s)=1$ and let
$C=\kappa n_0 \frac{V_0}{V}T_0\left(\frac{V_0}{V}\right)^{1/\kappa}$
to get
$ \epsilon(r_\rho,r_y)=Ce^{-s/2}$.
With $s=s_{\rho}+s_y$ where $s_\rho=r_\rho^2/R^2$ and $s_y=r_y^2/Y^2$.

We also get that
$$
E_{int}=C\int_{-aY}^{+aY} e^{-\frac{r_y^2}{2Y^2}}dr_y
\int_0^{bR} e^{-\frac{r_\rho^2}{2R^2}}2 \pi r_\rho dr_\rho \ .
$$
Using a change of variables to $s_y$ and $s_\rho$, so that
$dr_y=\frac{Y}{2\sqrt{s_y}}ds_y$ and
$2\pi r_\rho\, dr_\rho=R^2\pi ds_\rho$, the scaling
integration boundaries will be
$S_{yM} = a^2, \ \ $
$S_{\rho M} = b^2$
and we find
$$
E_{int}=C \pi R^2 Y \int_{-a^2}^{+a^2} e^{-s_y/2} \frac{ds_y}{\sqrt{s_y}}
\int_0^{b^2} e^{-s_\rho/2}\, ds_\rho \ .
$$
We can express the integrals as follows
$$
2\, I_A(b^2/2)\equiv \int_{0}^{b^2} e^{-s_\rho/2} ds_\rho=
2\int_{0}^{b^2/2} e^{-u} du
$$
and
\be
2\sqrt 2\,  I_B(a^2/2)\equiv
\int_{-a^2}^{a^2} e^{-s_y/2} \frac{ds_y}{\sqrt{s_y}} =
2\int_{0}^{a^2} e^{-s_y/2} \frac{ds_y}{\sqrt{s_y}} =
2\sqrt 2\int_{0}^{a^2/2} e^{-u} \frac{du}{\sqrt{u}} \ ,
\nonumber
\ee
where
\be
I_A(u) = 1-e^{-u} , \ \ \ {\rm and} \ \ \
I_B(u) = \sqrt \pi \Phi (\sqrt u)\ .
\nonumber
\ee

Thus
\ba
E_{int}&=&C \pi R^2 Y 2 I_A(b^2/2)2\sqrt 2 I_B(a^2/2)
\nonumber\\
&=& C \pi R^2 Y 4 \sqrt 2 \left(1{-}e^{-b^2/2}\right)
\sqrt \pi \ \Phi \left(a/\sqrt 2 \right). \
\ea

Similarly we can calculate the Kinetic energies,  $E_{kin}$,
for the rotation and radial and longitudinal expansions
as in
\cite{CWC2014}
(and see the Appendix).  
Then, using Eq. (\ref{Etot-1}) or (\ref{EC2}),
the total energy of the system is
\be
E_{tot} = E_{int} + E_{kin} \ ,
\ee
which we can use in the calculation of the weighted
vorticity.

In the present study we assumed an infinite system
with scaling gaussian density profile, Eq. (\ref{alternate}),
so that the integrals are evaluated up to infinity.

\begin{figure}[hb]  
\begin{center}
\resizebox{0.49\columnwidth}{!}
{\includegraphics{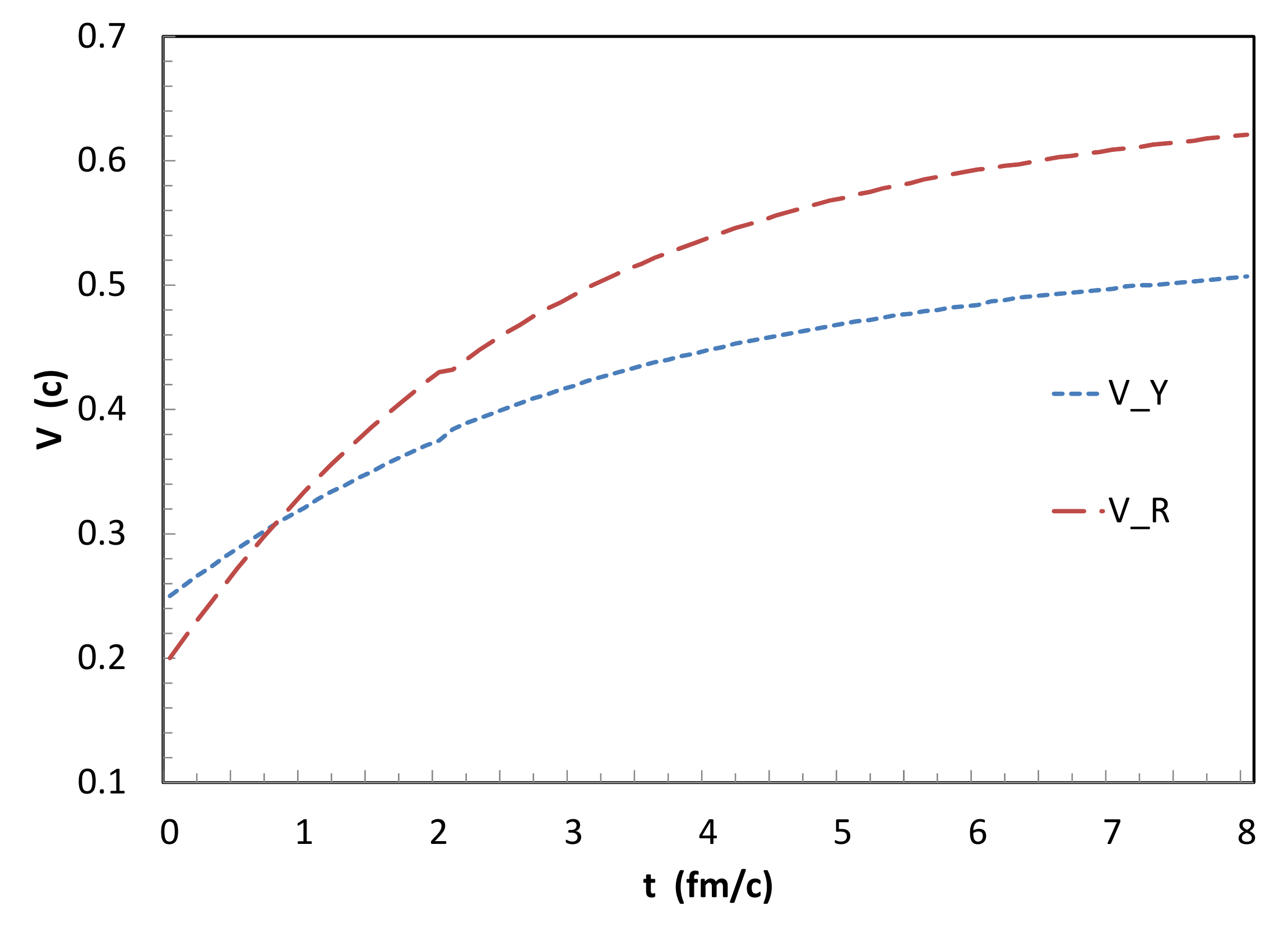}}
\hfill
\resizebox{0.49\columnwidth}{!}
{\includegraphics{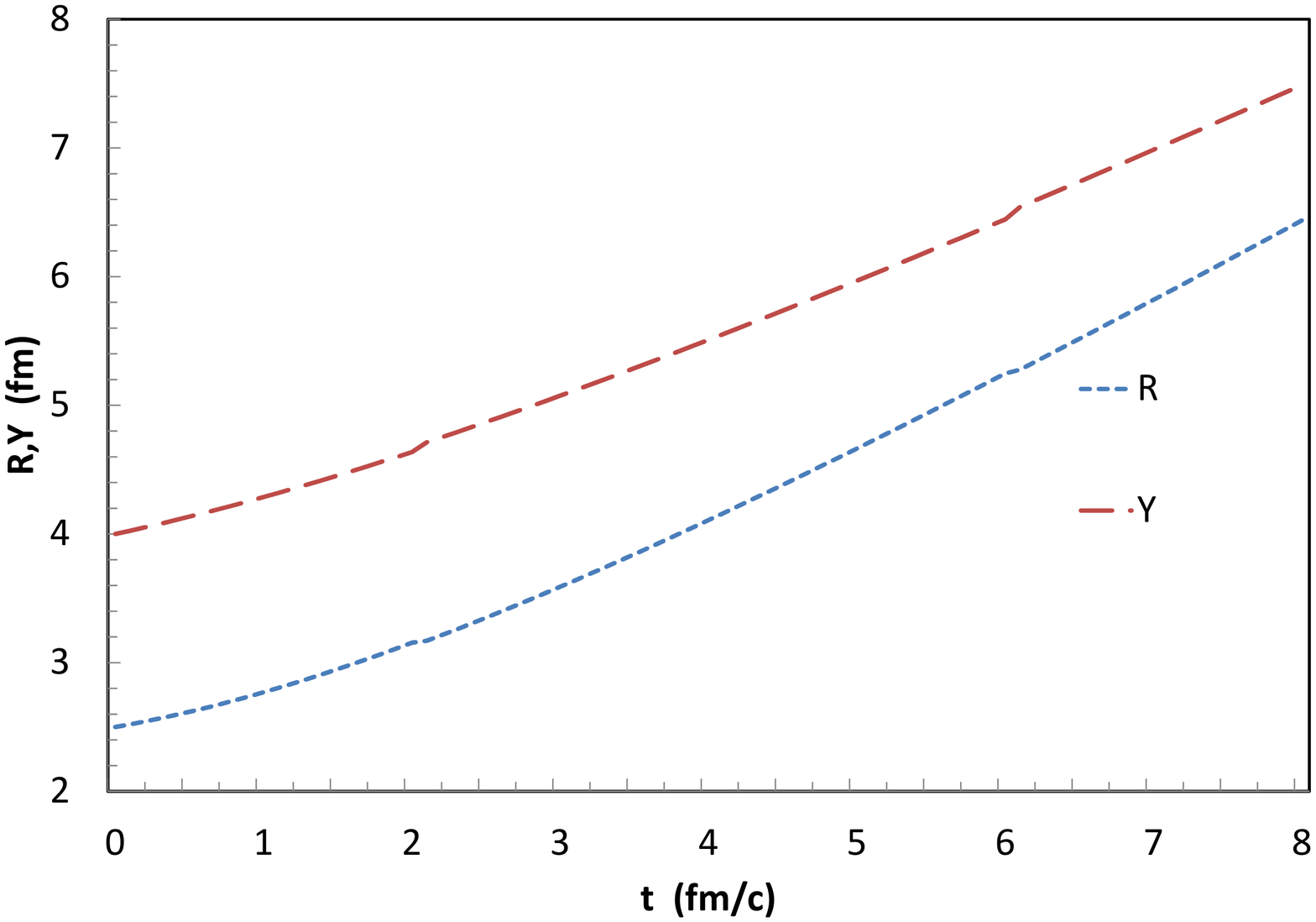}}
\caption{
(Color online)
Left: The time dependence of the velocity of expansion in the
transverse radial direction, $v_R$ and in the direction of
the axis of the rotation, $v_Y$, for the configuration presented in
Fig. \ref{F1-E-vs2-t}. The expansion velocity
is increasing in both directions. While in the axis direction
the velocity increases from 0.25 c to 0.5 c in 8 fm/c time, the
radial expansion increases faster.
Right: The time dependence of the Radial, $R$, and axis directed, $Y$,
size of the expanding system.
}
\label{F2-v-vs-t}
\end{center}
\end{figure}

\section{Results and Discussion}

We performed a set of calculations to study the applicability
of the model to heavy ion reactions. This presented non-relativistic model
leads to super-luminous velocities at late times and at the
external surface of the system.  We used a parametrization
where the peripheral energy density is cut off exponentially,
and took initial conditions such that the vast majority of the
system is in the non-relativistic applicable domain of the
model.

\begin{figure}[ht]  
\begin{center}
\resizebox{0.485\columnwidth}{!}
{\includegraphics{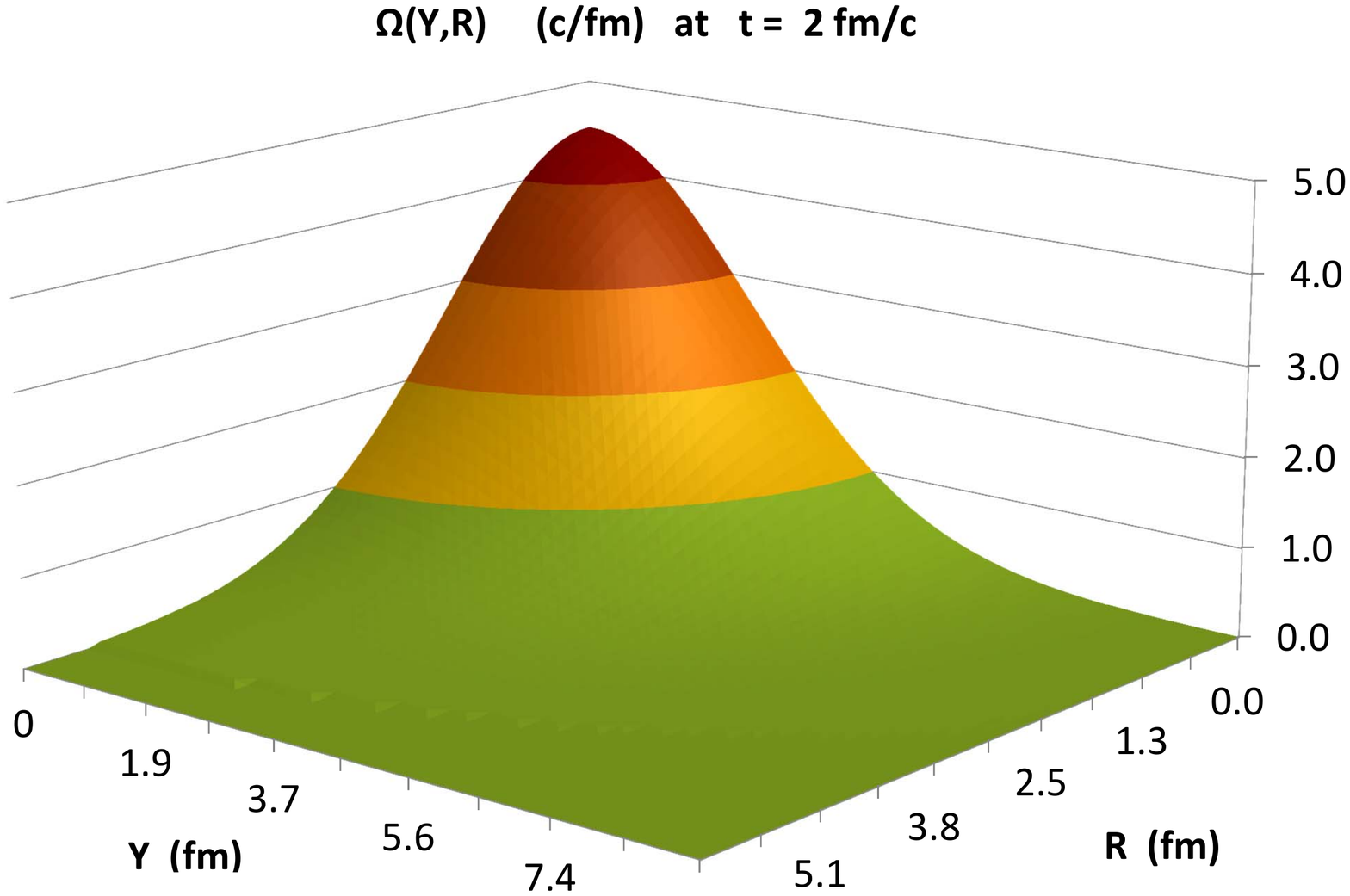}}
\hfill
\resizebox{0.485\columnwidth}{!}
{\includegraphics{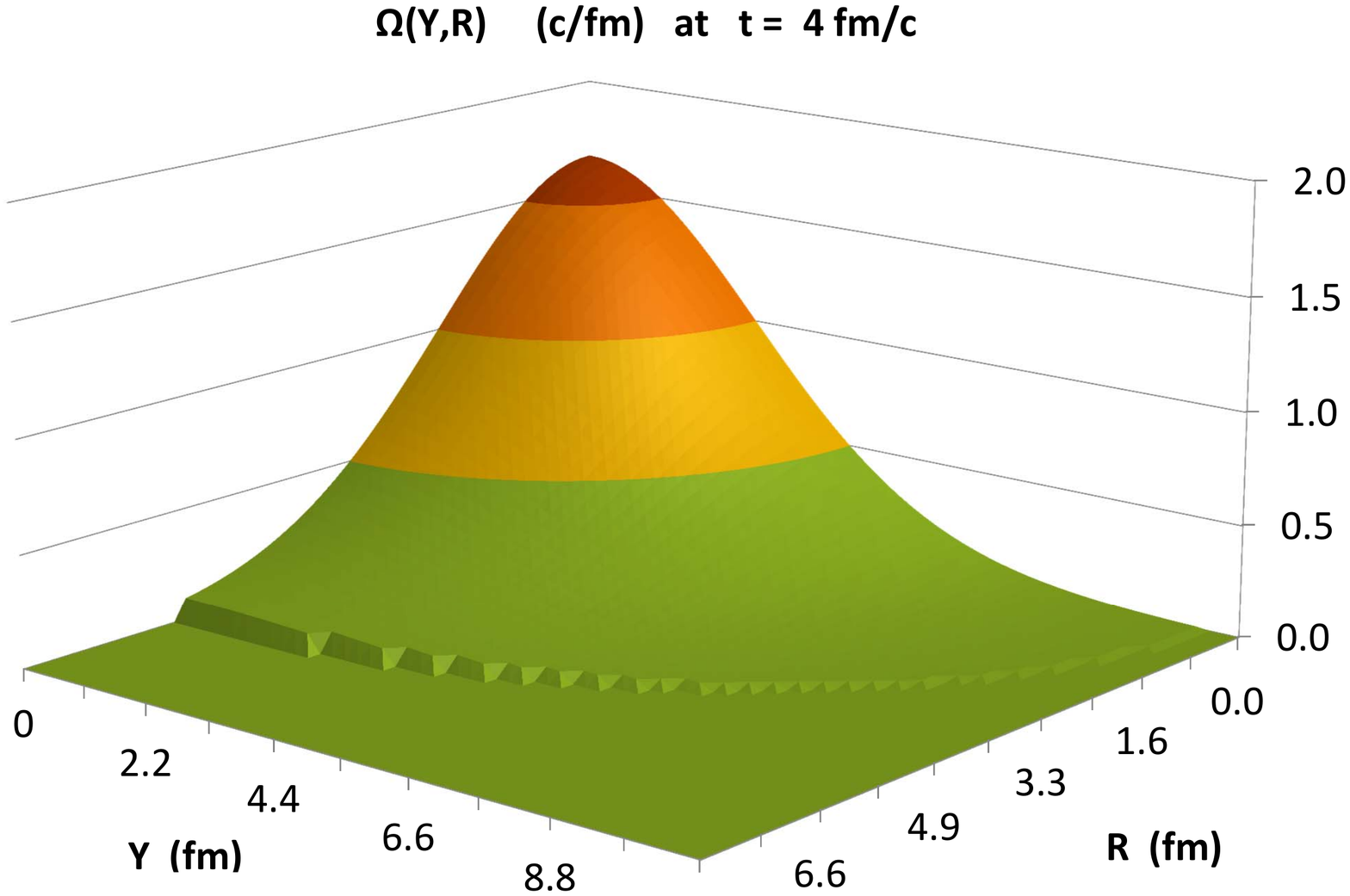}}\\
\resizebox{0.485\columnwidth}{!}
{\includegraphics{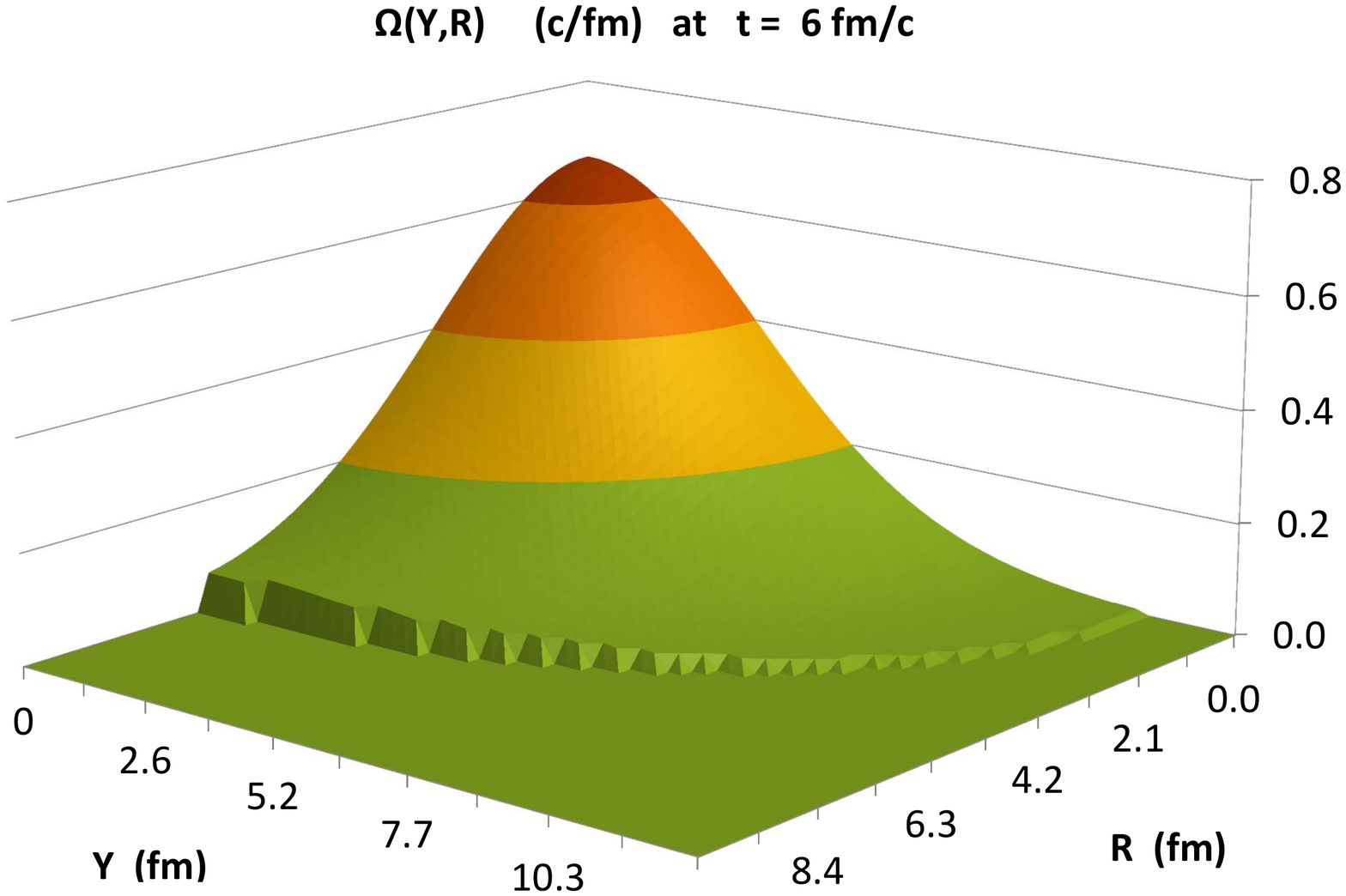}}
\hfill
\resizebox{0.485\columnwidth}{!}
{\includegraphics{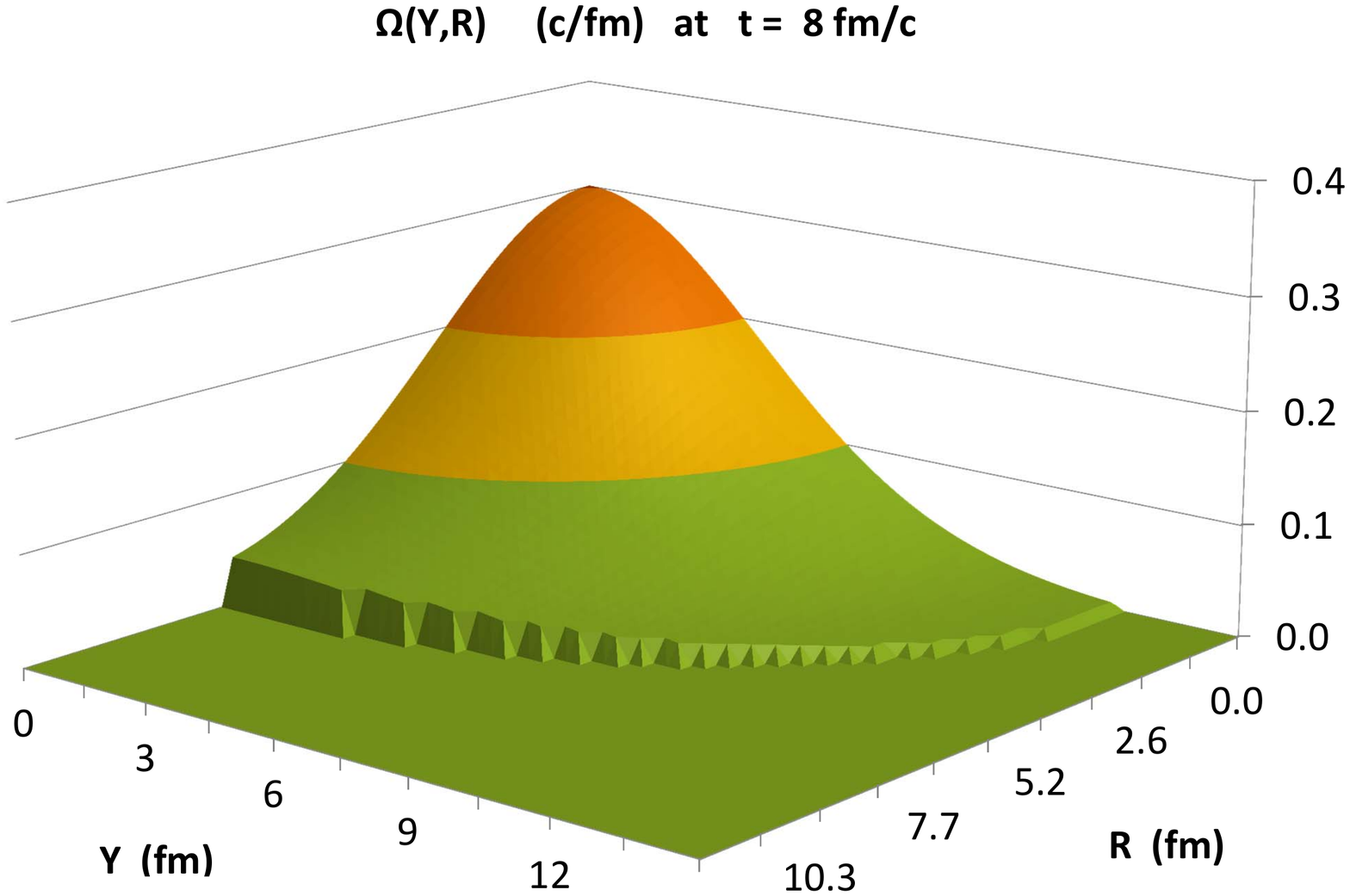}}
\caption{
(Color online)
The energy weighted vorticity in the classical rotating exact model
with Gaussian density profiles, with initial parameters as given in
Fig. \ref{F1-E-vs2-t}.
The figure
shows the configuration at
$t = 2$ fm/c, when
the mean radial (longitudinal) sizes and speeds are
3.16 fm (4.64 fm) and 0.43 c (0.38 c), and the
angular velocity is 0.06 c/fm.
The boundary is at the position where the velocity of matter reaches
the speed of light, c. This happens at
$Y_{max} > 9.0$ fm and $R_{max}= 5.68$ fm.
At $t = 4$ fm/c,
the mean radial (longitudinal) sizes and speeds are
4.11 fm (5.51 fm) and 0.54 c (0.45 c), and
$\omega = 0.04$ c/fm.
The $v = c$-boundary is at
$Y_{max} = 10.58$ fm and $R_{max}= 6.24$ fm.
At $t = 6$ fm/c,
the mean radial (longitudinal) sizes and speeds are
5.25 fm (6.45 fm) and 0.59 c (0.48 c), and
$\omega =  0.02$ c/fm.
The $v = c$-boundary is at
$Y_{max} = 11.34 $ fm and $R_{max}= 7.56$ fm.
At
$t = 8$ fm/c,
the mean radial (longitudinal) sizes and speeds are
6.44 fm (7.49 fm) and 0.62 c (0.51 c), and
$\omega =  0.02$ c/fm.
The $v = c$-boundary is at
$Y_{max} = 12.58$ fm and $R_{max}= 8.75$ fm.
}
\label{F5-t}
\end{center}
\end{figure}

A first series of calculations is presented in Figs.
\ref{F1-t}.

With the parameters as defined in Figs. \ref{F1-t} when we reach
$t = 8$ fm/c the surface speed reaches the speed of light
already when the Energy weighted vorticity drops to 40\% of the
top central value. Thus, a substantial amount of matter is
outside the range of physical applicability of the model.
The evaluation of polarization would not be realistic with
these sets of parameters.  Therefore we modified our initial
conditions such that the applicability of the non-relativistic model
holds up to the final, freeze out time of about 8 fm/c.

The time development of the change of the different forms
of energy are presented in Fig. \ref{F1-E-vs2-t}, for the modified initial
state, While the sizes $R$ and $Y$, and the expansion velocities
in these directions are shown in Fig. \ref{F2-v-vs-t}.

Now the total energy of the system is about 70\% of the previous example,
and the initial rotational energy is 60\% of the previous one.

We also performed another test series with this more compressed
initial state configuration
\ref{F5-t}.  
While up to $t = 6$ fm/c the majority of the energy weighted
vorticity is in the applicable domain (where the velocity does not
exceed the velocity of light), at $t = 8$ fm/c roughly 95\% of the
energy content is still in the applicability domain of the
non-relativistic exact model (see Fig. \ref{F5-t}).
We may estimate that about 50-70\%
of the initial energy of a peripheral collision will 
contribute to the expansion
of an symmetric solution of our participant system. 
Thus the model is applicable at
lower energies, FAIR and NICA, energies, while at the top energies
of RHIC or LHC the reliability of this model is qualitative, and
may provide estimates with 15 - 20 \% accuracy.

\section{Conclusions}

The effect of QGP formation on the directed flow and the arising
3rd flow component or antiflow was first observed in fluid dynamical
calculation at energies above 10 GeV per nucleon in Ref. \cite{BA94}.
The nuclear EoS has to satisfy strong constraints from
the observed Neutron and Hybrid Star masses \cite{RS92}
Spin-orbit interaction and the momentum dependence of the
nuclear interaction \cite{CF92}, influence the nuclear EoS and
developing rotation and polarization of the participant matter.
The nuclear EoS has a strong effect on the collective motion.
Transverse flow and collectivity was observed early both in
fluid dynamical, nuclear cascade and molecular dynamics models
\cite{AS91}.

In conclusion, the exact model can be well realized with
parameters extracted from detailed, high resolution, 3+1D
relativistic fluid dynamical model calculations with the
PICR code. It provides an estimate of the rate of decrease
of angular speed and rotational energy due to the
expansion in an explosively expanding system. This indicates
that the effects of rotation can be observable in case of
rapid freeze out and hadronization, although the Kelvin
Helmholtz Instability is not present in this model and this
reduces the rotation at later times.

This indicates that the presence of the KHI is essential
for an observable effect of the rotation, and thus the observation
of the rotation is strongly connected to the evolving
turbulent instability in low viscosity Quark-gluon plasma.

\bigskip

\section*{Acknowledgements}

Enlightening discussions with
Marcus Bleicher, Tam\'as Cs\"org\H o, Dariusz Miskowiec,
Horst St\"ocker, Sindre Velle and Dujuan Wang, are gratefully acknowledged.
\bigskip

\section*{References}


\newpage

\section{Appendix - Scaling of density distributions }
\label{A1}

Let us evaluate the baryon density, $n(s)$, and for simplicity
let us assume that in case 1A of Ref. \cite{CsNa13} the temperature
is constant, $\mathcal{T}(s) = 1$, then it follows that,
$\nu(s) = (N_B/V) \exp(-s/2)$, where $N_B = n_0\, V_0$.
Due to the exponential density profile, if $s$ is a sum of the
coordinates in two orthogonal directions, as $s = s_\rho = s_z$,
then n(s) separates into two multiplicative terms: $n_\rho(s_\rho)$
and $n_z(s_z)$.
For further simplifying the formalism, we can introduce a coordinate
change $s=2u$ for integrals of type $\int_0^{U/2} f(u/2)\, du$. Then $ds =
2\, du$ and  $2\int_0^{U/2} f(u/2)\, du = \int_0^{S} f(s)\, ds$.
This change will thus modify the upper limits of integration, and the
normalization by a factor of two.  These adjustments are included in the
final expressions in Appendices \ref{A1}-\ref{A4}.

The baryon density distribution is then
$$
n(r_\rho, r_y) = N_B \frac{C_n}{V} e^{-r_\rho^2/R^2} e^{-r_y^2/Y^2}
$$
where $C_n$ is a normalization constant, which will be determined later.
The normalization can be performed up to a finite size, $R$ and $Y$, or
up to infinity.
\ba
\int\!\!\! \int\! n(r_\rho, r_y) &=&  N_B \frac{C_n}{V}
\nonumber \\
2\pi
 &\times& \int_0^{R/\infty} e^{-r_\rho^2/R^2} r_\rho dr_\rho
\nonumber \\
 &\times& \int_{-Y/\infty}^{Y/\infty} e^{-r_y^2/Y^2} dr_y\ ,
\ea
here the first integral up to infinity gives $\Gamma(1) R^2/2$, while the
second one $\Gamma(0.5) Y = \sqrt{\pi} Y$.
In $x,\, y,\, z$ coordinates this is:
\ba
N_B \frac{C_n}{V} &\times& \left(
\int_{-\infty}^{\infty} e^{-r_x^2/\dot{R}^2} dr_x \right)^3 =
\nonumber \\
N_B \frac{C_n}{V} \left( \sqrt{\pi} \dot{R} \right)^3 &=& N_B \times {\rm const.}
\ea
Or in cylindrical coordinates
\ba
N_B \frac{C_n}{V} &\times& \pi \Gamma(1) R^2 \times \sqrt{\pi} Y =
N_B \frac{C_n}{V} \pi^{3/2} R^2 Y \ ,
\ea
which is the same. The integrals were evaluated up to limits
in infinity. If we perform the definite integrals up to a finite
limit, we get similar scaling behaviour. Let us now change the variables
to scaling variables introduced in Ref. \cite{CsNa13}, but in
cylindrical coordinates.
$$
n(s_\rho, s_y) = N_B \frac{C_n}{V}\, e^{-s_\rho/2}\, e^{-s_y/2} .
$$
Now using the relations
$ 2 \pi r_\rho dr_\rho = \pi R^2 ds_\rho$ and
$ dr_y = \frac{Y}{2 \sqrt{s_y}} ds_y$  we get
\ba
\int\!\!\! \int\! n(s_\rho, s_y) &=&  N_B \frac{C_n}{V}
\nonumber \\
 &\times& \pi R^2\!\! \int_0^1\!\! e^{-s_\rho/2}\,  ds_\rho
  \times        Y\!\! \int_0^1\!\! e^{-s_y/2}\, \frac{ds_y}{ \sqrt{s_y}} =
\nonumber \\
 N_B {C_n} \frac{\pi R^2Y}{V}
 &\times& 2\!\! \int_0^{0.5}\!\! e^{-u}\,  du
  \times  \sqrt{2}\!\! \int_0^{0.5}\!\! e^{-u}\, \frac{du}{\sqrt{u}}
\nonumber \\
&=& N_B C_n \frac{\pi R^2 Y}{V}\ 2\, I_A(0.5)\ \sqrt{2}\, I_B(0.5)
\ea
This should be equal to $N_B$ thus the normalization constant is
$$
C_n = 1 \left/ \left[  2\sqrt{2}\, I_A(0.5)\, I_B(0.5) \right] \right. .
$$
Here  $I_A(0.5)$ and  $I_B(0.5)$ are constants, which do not change
during the scaling evolution, when the density profile remains the
same. At infinity
$I_A(\infty) = \Gamma(1)$ while
$I_B(\infty) = \Gamma(0.5) = \sqrt{\pi}$, but at different integration
limits the ratio of the two integrals will be different \cite{IGF}:
\ba
I_A(u) &=& 1 - \exp(-u)
\nonumber \\
I_B(u) &=& \sqrt{\pi}\, \Phi(\sqrt{u}\,) \ ,
\ea
where
\be
\Phi(u)= {\rm erf}(u) \equiv
\frac{2}{\sqrt{\pi}} \int_0^u \exp(-x^2)\, dx \ .
\ee

\section{Appendix - The Moment of Inertia}
\label{A2}

Consider a body with scaling expansion, and with solid
body rotation (i.e. the angular velocity is uniform for the
whole body, $\omega = \omega(t)$ but it does not depend on the
spatial coordinates.
Let us denote the moment of inertia
with $\Theta$,
\be
\Theta = \int m\,  n(r)\,  r^2 \,  d^3r \ .
\ee
Then the angular momentum and the rotational energy are
\be
L = \Theta \omega \ \ \ {\rm and} \ \ \
E = \frac{1}{2} \Theta \omega^2 \ .
\ee
Now we assume that our system has no external torque, and all internal
forces are radial, so the angular momentum must be conserved, during the
scaling expansion driven by the pressure gradient which is
radial in a cylindrically symmetric system. Thus, the angular velocity
is not directly influenced by the dynamics, just via the angular momentum
conservation. From
$
\dot{L} = 0  \ ,
$
it follows that
$$
\dot{\Theta} \omega = - \Theta \dot{\omega} \ \ \
{\rm or} \ \ \
\dot{\omega} = - \omega \frac{\dot\Theta}{\Theta} \ .
$$
Thus the change of the angular velocity is a direct consequence of the
change of the moment of inertia $\Theta$, while  $\Theta$ is
proportional with the square of the radius of the system
in a scaling expansion where the density profile remains the
same during the expansion. Consequently
$$
\dot{\omega}\  =\ - \omega \frac{\dot\Theta}{\Theta}\ =\
- \omega \frac{\dot{R}^2}{R^2} \ \ \ {\rm and} \ \ \
\omega \ =\ \omega_0 \frac{R_0}{R}\ .
$$
We still have to evaluate the moment of inertia accurately to provide
precisely the energy of rotation. Thus,
using the scaling variables
\ba
\Theta &=& m N_B \frac{\pi R^2\, Y}{V}
     R^2 C_n\!\! \int_0^1\!\! e^{-s_\rho/2} s_\rho  ds_\rho
 \times\!\!  \int_0^1\!\! e^{-s_y/2} \frac{ds_y}{ \sqrt{s_y}}
\nonumber \\
&=& m\, N_B\, R^2\ C_n\ 4\, I_C(0.5) \, \sqrt{2}\, I_B(0.5) \ ,
\ea
where
$
 I_C(u) = 1 - (1+u) \exp(-u) \ .
$

As before these integrals do not change during the scaling expansion,
on the other hand the volume and the moment of inertia have
different coefficients in the energy expression.
As a consequence the kinetic energy of the rotation is
\ba
E_{kin} &=& \frac{1}{2} \Theta \omega^2 =
 \frac{1}{2}  m\, N_B\ C_n 4\sqrt{2}\, I_C(0.5)\, I_B(0.5)\,  R^2 \omega^2
\nonumber \\
   &=&   \frac{1}{2} \alpha^2\, m\, N_B\   R^2 \omega^2\ .
\ea
Here we have introduced the constant
\be
\alpha^2 \equiv
4\sqrt{2}\,C_n I_B({\small \frac{1}{2}}) I_C({\small \frac{1}{2}}) ,
\ee
that can be used in the main course of the work.

\section{Appendix - Kinetic energy of radial expansion}
\label{A3}

The radial velocity is given by $v_\rho = (\dot{R}/R) r_\rho$ and consequently
 $v_\rho^2 = \dot{R}^2 s_\rho$. Thus the kinetic energy of radial expansion is
\ba
E_{kin} &=& \frac{m N_B}{2} \frac{\pi R^2 Y}{V} \dot{R}^2 C_n
\int_0^1\!\!\! e^{-s_\rho/2} s_\rho ds_\rho
\int_0^1\!\!\! e^{-s_y/2} \frac{ds_y}{\sqrt{s_y}}
\nonumber \\
 &=& \frac{m N_B}{2} C_n  4\sqrt{2}\,I_C(0.5)\, I_B(0.5)\ \dot{R}^2
\nonumber \\
  &=& \frac{m N_B}{2}\, \alpha^2\ \dot{R}^2
\ea
\bigskip

\section{Appendix - Kinetic energy of longitudinal expansion}
\label{A4}

The longitudinal velocity is given by $v_y = (\dot{Y}/Y) r_y$ and
consequently
$v_y^2 = \dot{Y}^2 s_y$. Thus the kinetic energy of longitudinal expansion is
\ba
E_{kin} &=& \frac{m N_B}{2} \frac{\pi R^2 Y}{V} \dot{Y}^2  C_n\!\!
\int_0^1\!\!\! e^{-s_\rho/2} ds_\rho
\int_0^1\!\!\! e^{-s_y/2} \sqrt{s_y} ds_y
\nonumber \\
 &=& \frac{m N_B}{2} C_n\ 4\sqrt{2}\, I_A(0.5)\,I_D(0.5)\ \dot{Y}^2 \ ,
\ea
where
$
I_D(u) = \frac{\sqrt{\pi}}{2} \Phi(\sqrt{u}) - \sqrt{u} e^{-u}.
$
Here we can introduce the constant
\be
 \beta^2 \equiv
4\sqrt{2}\,C_n I_A({\small \frac{1}{2}}) I_D({\small \frac{1}{2}}) ,
\ee
which will be used in the calculation.

\section{Appendix - Not realizable analytic solution.}
\label{A5}

One may find a solution for the dynamical evolution of
$R(t)$ and $Y(t)$ based on Eq. \ref{EC1},
with simplifying the problem to a
singe, first order differential equation in a similar way as it is
done in Ref. \cite{Akkelin01}.
Let us introduce a parametric function, $\phi(t)$,
\ba
\alpha R(t) &=& U(t) \sin \phi(t) \nonumber \\
\beta  Y(t) &=& U(t) \cos \phi(t)  \ ,
\label{RZU}
\ea
satisfying Eq. (\ref{eu}).
Now inserting Eqs. (\ref{RZU}) into Eq. (\ref{EC1}),
and noticing that
$\alpha^2 \dot{R}^2 + \beta^2 \dot{Y}^2 = \dot{U}^2 + U^2 \dot{\phi}^2$
we get the following first order differential equation for
$\phi$:
\be
\dot{\phi}^2 =
\frac{1}{U^2(t)}  \!\!\! \left[ \
{\cal F} - \,
\dot{U}^2(t) - \,
\frac{\alpha^4 W}{U^2(t) \sin^2 \phi} - \,
\frac{(\alpha^2+\beta^2) (\alpha^2\, \beta)^\gamma Q }
{( \pi U^3(t) \sin^2\phi \cos \phi)^\gamma} \
 \right] .
\label{eqPhi}
\ee
\smallskip

The initial value of the variable $\phi$ is chosen such that Eq.
(\ref{RZU}) is satisfied for $U(t_0),\ \phi(t_0)$.

The problem with this solution is that Eq. (\ref{eqPhi})
describes the square of $\dot{\phi}$ and in a realistic situation it
is not trivial to find the sign of the r.h.s. of the dynamical equation
for  $\dot{\phi}$. This sign alternates.


\begin{thebibliography}{00}


\bibitem{xnwang}
 J.~H.~Gao, Z.~T.~Liang, S.~Pu, Q.~Wang
  and X.~N.~Wang, Phys.\ Rev.\ Lett.\  {\bf 109}, 232301 (2012).		

\bibitem{CMW13}
  L.P. Csernai, V.K. Magas, D.J.~Wang,  Phys. Rev. C {\bf 87},
034906 (2013).

\bibitem{CKM}
  L.P.~Csernai, J.I.~Kapusta, L.D.~{McLerran},
  Phys.~Rev.~Lett. {\bf 97},  152303 (2006).

\bibitem{hydro1}
  L.P. Csernai, V.K. Magas, H. St\"ocker, and D.D. Strottman,
  Phys. Rev. C {\bf 84},  024914 (2011).

\bibitem{hydro2}
  L.P. Csernai, D.D. Strottman and Cs. Anderlik,
  Phys. Rev. C {\bf 85}, 054901 (2012).

\bibitem{WNC13}
  D.J. Wang, Z. N\'eda, and L.P. Csernai,
  Phys. Rev. C {\bf 87}, 024908 (2013)

  \bibitem{CK85}
L.P. Csernai, and J.I. Kapusta,
Phys. Rev. D 31, 2795 (1985).



\bibitem{CWC2014}
  L.P. Csernai, D.J. Wang  and T. Cs\"org\H o,
  Phys. Rev. C {\bf 90}, 024901 (2014).
  
\bibitem{Hatta}
Y. Hatta, J. Noronha, Bo-Wen Xiao,
Phys. Rev. D 89, 051702 (2014).

\bibitem{CsNa13}
T. Cs\"org\H o and M.I. Nagy,
Phys. Rev. C {\bf 89}, 044901 (2014).

\bibitem{Stoecker_handbook}
Horst St\"ocker: {\it Taschenbuch Der Physik},
(Harri Deutsch, 2000), 1.3.2/6d.

\bibitem{IGF}
  M. Abramowitz, and I.A. Stegun: {\it Handbook of mathematical functions}
(Dover, New York, 1965) 6.5.2;
I.S. Gradstein, and I.M. Ryzhik: {\it Table of Integrals ...},
(Academic Press, 1994)
3.321/2., 3.361/1., 3.381/1., 8.250/1., 8.251/1., 8.350/1., 8.354/1.

\bibitem{Akkelin01}
S.V. Akkelin, T. Cs\"org\H{o}, B. Luk\'acs, Yu. M. Sinyukov, and
M. Weiner, Phys. Lett. B {\bf 505} (2001) 64-70.

\bibitem{RungeKutta} W.E. Boyce and R.C. DiParma: {\it
Elementary Differential Equations and Boundary Value Problems},
(Wiley, 1997).

\bibitem{BA94}
L.V. Bravina, N.S. Amelin, L.P. Csernai, P. L\'evai, and D. Strottman,
Nuclear Physics A 566, 461-464 (1994).

\bibitem{RS92}
A. Rosenhauer, E.F. Staubo, L.P. Csernai, T. Overgard, and E. Ostgaard,
Nucl. Phys. A 540, 630 (1992).

\bibitem{CF92}
L.P. Csernai, G. Fai, C. Gale, and E. Osnes,
Phys. Rev. C 46, 736 (1992).

\bibitem{AS91}
N.S. Amelin, E.F. Staubo, L.P. Csernai, V.D. Toneev, K.K. Gudima, and
D. Strottman,
Phys. Rev. Lett. 67, 1523 (1991).

\end{thebibliography}
\end{document}